\documentclass{sig-alternate} 
\usepackage{amsmath,bm}
\usepackage{graphicx,subfigure}
\usepackage{multirow}
\usepackage{tikz}
\usetikzlibrary{arrows}
\tikzstyle{block}=[draw opacity=0.7,line width=1.4cm]

\title{Conquering the rating bound problem in neighborhood-based collaborative filtering: a function recovery approach}
\numberofauthors{4}
\author{
\alignauthor
Junming Huang\\
       \affaddr{Institute of Computing Technology}\\
       \affaddr{Chinese Academy of Sciences}\\
       \affaddr{Beijing, China}\\
       \email{mail@junminghuang.com}
\alignauthor
Xue-Qi Cheng\\
       \affaddr{Institute of Computing Technology}\\
       \affaddr{Chinese Academy of Sciences}\\
       \affaddr{Beijing, China}\\
       \email{cxq@ict.ac.cn}
\alignauthor
Hua-Wei Shen\\
       \affaddr{Institute of Computing Technology}\\
       \affaddr{Chinese Academy of Sciences}\\
       \affaddr{Beijing, China}\\
       \email{shenhuawei@ict.ac.cn}
\and  
\alignauthor
Xiaoming Sun\\
       \affaddr{Institute of Computing Technology}\\
       \affaddr{Chinese Academy of Sciences}\\
       \affaddr{Beijing, China}\\
       \email{sunxiaoming@ict.ac.cn}
\alignauthor
Tao Zhou\\
       \affaddr{Web Sciences Center}\\
       \affaddr{University of Electronic Science and Technology of China}\\
       \affaddr{Chengdu, China}\\
       \email{zhutou@ustc.edu}
\alignauthor
Xiaolong Jin\\
       \affaddr{Institute of Computing Technology}\\
       \affaddr{Chinese Academy of Sciences}\\
       \affaddr{Beijing, China}\\
       \email{jinxiaolong@ict.ac.cn}
}

\begin{document}

\maketitle

\begin{abstract}
As an important tool for information filtering in the era of socialized web, recommender systems have witnessed rapid development in the last decade. As benefited from the better interpretability, neighborhood-based collaborative filtering techniques, such as item-based collaborative filtering adopted by Amazon, have gained a great success in many practical recommender systems. However, the neighborhood-based collaborative filtering method suffers from the rating bound problem, i.e., the rating on a target item that this method estimates is bounded by the observed ratings of its all neighboring items. Therefore, it cannot accurately estimate the unobserved rating on a target item, if its ground truth rating is actually higher (lower) than the highest (lowest) rating over all items in its neighborhood. In this paper, we address this problem by formalizing rating estimation as a task of recovering a scalar rating function. With a linearity assumption, we infer all the ratings by optimizing the low-order norm, e.g., the $l_\frac{1}{2}$-norm, of the second derivative of the target scalar function, while remaining its observed ratings unchanged. Experimental results on three real datasets, namely Douban, Goodreads and MovieLens, demonstrate that the proposed approach can well overcome the rating bound problem. Particularly, it can significantly improve the accuracy of rating estimation by $37\%$ than the conventional neighborhood-based methods.
\end{abstract}


\section{Introduction}

With the explosion of web information in the last decade, it becomes more and more difficult for individuals to discover interesting information from massive web resources. To solve the information overload problem that has attracted lots of attention from both academic and industrial communities, various personalized recommender systems have been developed to help a user automatically find their interested information. Mainstream approaches include kNN collaborative filtering \cite{item-based-collaborative-filtering}, latent factor models \cite{svd-plus-plus, matrix-factoriaztion-1, matrix-factoriaztion-survey}, resource projection \cite{resource-projection}, restricted Boltzmann machine \cite{{restricted-boltzmann-machine}}, etc. Although in recent years recommender systems play a more and more important role in online commerce and sharing services such as Amazon, Netflix and YouTube, there still exists much room for a recommender system to improve its accuracy. A bunch of problems including sparsity, cold start and diversity remain as grand challenges in the open literature \cite{collaborative-filtering-survey, collaborative-filtering-survey-2}.

The two major schools, neighborhood-based methods and latent factor models, are facing their own difficulties respectively. The family of latent factor models, originating from matrix factorization and making great progress by incorporating with probabilistic graphical models \cite{probabilistic-matrix-factorization} and compressed sensing \cite{matrix-factoriaztion-cs} recently, gains success in accurately completing missing ratings. However, its lack of interpretability may limit its application in practice, since interpretability plays a critical role in practice to affect users' experience \cite{explainable}. The conventional neighborhood-based approaches like kNN collaborative filtering techniques produce estimations in a way much easier to explain clearly, but meanwhile suffer from the \emph{rating bound problem}. Take the item-based collaborative filtering as example. A neighborhood-based approach estimates an unobserved rating with the weighted average of ratings on similar items called neighbors, and therefore the estimation is bounded by observed neighbor's ratings. However, an item a user loves or hates \emph{the most} is usually rated higher or lower than all the neighbor items, and therefore has no chance to be correctly predicted by a neighborhood-based approach. Similarly, the rating bound problem is also a big challenge to the user-based collaborative filtering technique. 

The rating bound problem will not be problematic if the actual bounds are fully observed, i.e., the items rated higher/lower than all other items at least in a local range. Unfortunately, in many practical scenarios, those ratings are not always fully observed, since, for example, a movie fan is lazy to label all her favorite movies on each of a dozen movie websites she registers on, or a reader is reluctant to post a rating to an extremely boring book. Lacking observations of actual rating bounds, a neighborhood-based approach like kNN collaborative filtering technique seriously suffers from the rating bound problem because of its results being bounded by incorrect bounds. By empirical analysis, up to $15\%$ estimation tasks suffer from the rating bound problem. Nearly a half of estimation errors of kNN collaborative filtering owe to incorrect estimations on those items. This implies that it is a much more difficult job to accurately recover an unobserved rating with the rating bound problem than recovering an unobserved rating without the problem.

Besides the help to accurately recover missing ratings, the items which are rated higher or lower than all other items have their own values. Such items are called a user's \emph{interest centers}. A positive interest center of a user, i.e., an item she rater higher than all neighbor items, is an item she loves the most, such as five favorite restaurants in the city or most beloved movies in a library. To correctly discover and recommend those items (if unseen yet) makes the user enjoyable and trust the system. Symmetrically, a user might have several negative interest centers that she dislikes or hates, e.g., a soft music fan might be unhappy to find a heavy metal rock album in a recommendation list. A recommender system should try its best to avoid recommending those disliked items.

In a word, it is critical for a recommender system to solve the rating bound problem, not only for more accurately predicting the unobserved ratings, but also for better sketching a user's interest map and improving user experience. Latent factor models might be helpful to reduce the pain caused by the rating bound problem. Nevertheless, due to the practical importance to explain to users why the recommendation list is produces, as well as the difficulty in explaining matrix factorization results, we attempt to solve the problem in the line of neighborhood-based methods, which provides more explainable results than a latent factor model.



To address the rating bound problem, we view the task to estimate unobserved ratings in a recommender system as a job of function recovery. Given an item-item network built in the same way as in a standard neighborhood-based method, for each user $u$, a scalar function $r_{u}(\cdot)$ is defined on the network to map any item to a rating. A recommender system is required to recover the whole function $r_u(\cdot)$ as accurate as possible, based on the partial observation of the function value on a few items. With a practically verified prior knowledge that such a scalar function is linear on most items, i.e. its second derivative vanishes on most items, we develop an effective method to recover the function by minimizing the number of items with non-zero second derivatives (for simplicity, we denote \emph{sources} for items with non-zero values of the second derivative of a scalar function in the rest of the paper). Empirical practice supports that our approach effectively improves the performance when predicting items with the rating bound problem.

The major contributions in this paper are listed below,

\begin{itemize}
  \item We study the rating bound problem that the conventional neighborhood-based approaches suffer from.
  \item To solve the problem, we introduce a scalar function view to consider a recommender system algorithm as a scalar function recovery task based on partial observations.
  \item We propose an approach that minimizes the $l_\frac{1}{2}$-norm of the second derivative of a scalar function to recover it. The approach is validated effective with empirical experiments.
\end{itemize}

The rest of the paper is organized as follows. Section \ref{section:model} introduces our view to the recommender systems and Section \ref{section:inference} describes our approach to solve the rating bound problem, which is validated in Section \ref{section:experiments}. Section \ref{section:backgrounds} reviews the recent progress in recommender system research. Section \ref{section:conclusion} concludes the paper.

\section{Scalar functions recovery}\label{section:model}

In this section we introduce our view of recommender systems as a task of scalar functions recovery, and how different approaches leverage a property of the functions for inference.

From a function perspective, the process to complete missing ratings in a recommender system can be considered as a job of function recovery. For each user $u$, a scalar function $r_u(\cdot)$ is defined to map any item $i$ (e.g, book, music, movie, product, celebrity, etc.) to a real or integer rating $r_u(i)$. A recommender system is expected to recover the whole function $r_u(\cdot)$ for each user based on partial observation of the function value. Obviously this is impossible unless prior knowledge or additional evidence is provided.

Different prior knowledge or assumptions result in different approaches. For example, latent factor models assume that the explicit form of a scalar function is a linear combination. Each item is represented with a vector where each element corresponds to its ``quality score'' in a certain feature, and a set of ``interest'' weights is defined for each user to add up those scores to make a rating. Differently, neighborhood-based approaches do not assume the explicit for the a scalar function, but instead assume that the shape of the concerned function is ``smooth'' on an item-item network, and therefore can be fulfilled with interpolation. In the scope of this paper we extend the line of neighborhood-based approaches because of its ease in explanation. 

A widely believed assumption, usually called the similarity assumption, tells that if two items were rated similarly in the past, they will be rated similarly in the future. The assumption provides the basis to build the collaborative filtering technique, which estimates an unobserved rating with the weighted average of ratings on similar items. An equivalent description to the assumption tells the linearity of a scalar function defined on an item-item network where nodes are items and edges
describe the similarity among items \footnote{In the scope of this paper we discuss the item-based collaborative filtering. A similar conclusion to its user-based version follows by symmetry.}. We introduce a \emph{linearity assumption} of a scalar function that its second derivative vanishes on most items, i.e., the following equation holds for most items $i$,

\begin{equation}\label{equ:2nd-derivative-vanish}
\nabla^2 r_u(i) = 0,
\end{equation}
where $\nabla^2$ denotes a discrete second derivative operator. The linearity assumption helps us recover the whole function after observing part of function values.

\subsection{Equivalence explanation}
We first explain why the similarity assumption and the linearity assumption are equivalent. Let us recall the well-known Resnick equation applied in kNN collaborative filtering to predict an unknown rating $r_u(i)$ as follows\footnote{In the scope of this paper we use Equation (\ref{equ:resnick}) to describe kNN collaborative filtering. There are several forms of kNN collaborative filtering, for example
$\hat{r}_u(i)=\bar{r}(i) + \frac{\sum_{j\in N(i)} w(i,j) (r_u(j)-\bar{r}(j))}{\sum_{j\in N(i)} w(i,j)}$, where $\bar{r}(i)$ denotes the global average rating of item $i$. The discussion in this paper can be easily extended to the above form, with a simple preprocess to replace all $r_u(j)$ with $r_u(j)-\bar{r}(j)$. Extensions to other forms are similar.},

\begin{equation}\label{equ:resnick}
\hat{r}_u(i) = \frac{\sum_{j\in N(i)} w(i,j) r_u(j)}{\sum_{j\in N(i)} w(i,j)},
\end{equation}
where $w(i,j)$ is the weight on the edge between item $i$ and item $j$, i.e., the similarity between them. The weighted average is calculated among items in $N(i)$, the neighbor set of item $i$. We move the RHS to the left and have

$$
r_u(i) - \sum_{j\in N(i)}\frac{w(i,j)}{\sum_{j\in N(i)} w(i,j)} r_u(j) = 0.
$$

Labeling all items with integers $1,2,\dots,n$, we can rewrite the above equation in a matrix form,

$$
(I - D^{-1}W)\bm{R_u} = 0,
$$
where $\bm{R_u}$ is a vector consisting of $r_u(\cdot)$ values, $I_{n\times n}$ is an identity matrix, $W_{n\times n}$ is the weight matrix with element $W_{ij}=w(i,j)$, and the diagonal matrix $D_{n\times n}$ is defined as

\begin{equation}\notag
D_{ij} = \begin{cases}
           \enspace \sum_l w(i,l) & i=j, \\
           \enspace 0 & i\neq j.
         \end{cases}
\end{equation}

Notice that the Laplacian matrix $L = I - D^{-1}W$ is the negative of $\nabla^2$, which is the discrete second derivative operator. The above equation is exactly an second-order ordinary differential equation 

$$
L\bm{R_u} = -\nabla^2 \bm{R_u} = 0,
$$
as we mentioned in Equation (\ref{equ:2nd-derivative-vanish}). The connection was firstly introduced in \cite{heat-conduction}.

\subsection{Examining the linearity assumption}

To support the linearity assumption, we empirically examine whether the second derivative of any user's rating function vanishes on most items. In three real datasets (dataset details described in Section \ref{subection:dataset}), we collect such examples where the second derivative $\nabla^2 r_u(i)$ can be directly calculated, i.e., the ratings a user $u$ posts to an item $i$ and all its neighbor items in $N(i)$ are completely observed \footnote{Practically, since the fully observed examples are too few to build a solid statistics conclusion, we search for examples $r_u(i)$ such that the user $u$ rates the item $i$ and no less than $90\%$ neighbor items}. We further require that no less than $5$ ratings are observed on neighbor items, otherwise the calculation of $\nabla^2 r_u(i)$ might be unreliable.

\begin{figure}
\center
\includegraphics[width = 0.45 \textwidth]{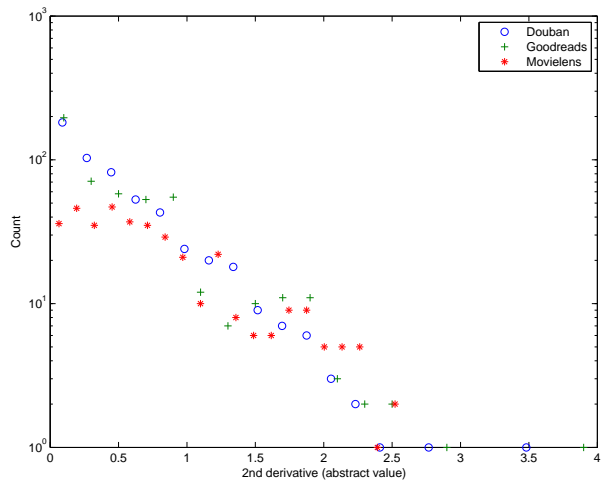}
\caption{Number of examples that a user's rating on an item drifts away from an estimator which is the weighted average among neighbor items. Examples are not presented if the concerned item has less than $5$ neighbors or less than $90\%$ neighbors rated by the same user. In a major of examples the drift is quite small, supporting the linearity assumption that the second derivative of any user's rating function vanishes on most items.}
\label{fig:exam.assumption}
\end{figure}

In each dataset, there exist hundreds of user-item pairs where the second derivative can be calculated according to observed ratings. The statistics of the obtained second derivatives are report in Fig. \ref{fig:exam.assumption}. As shown in the figure, on most examples the calculated values of second derivative are almost zero, and few examples have a second derivative far away from zero. The number of those examples decreases exponentially with the distance from zero (Note that the vertical axis is labeled in logarithm). The results confirm the linearity assumption on the shape of a user's rating function, which later helps us accurately recover the whole rating function based on an observed part of ratings.

\subsection{Leveraging the linearity assumption}\label{subsection:leverage-the-assumption}

\begin{figure}
    \centering
    \subfigure[kNN]{
        \label{fig:edge.usage.knn}
        \begin{tikzpicture}[node distance=15mm, thick]
        \tikzstyle{every node}=[minimum size=7mm]

        \node [fill=black!30!white ,draw=black,rectangle] (A) {5};
        \node [fill=white,draw=black,circle] (B) [right of=A] {5};
        \node [fill=black!30!white ,draw=black,rectangle] (C) [below of=A] {3};
        \node [fill=white,draw=black,circle] (D) [below of=B] {3};

        \draw[->] ([yshift=0.1 cm]A.east) -- ([yshift=0.1 cm]B.west) ;
        \draw[->,dashed] ([yshift=-0.1 cm]B.west) -- ([yshift=-0.1 cm]A.east) ;

        \draw[->] ([yshift=0.1 cm]C.east) -- ([yshift=0.1 cm]D.west) ;
        \draw[->,dashed] ([yshift=-0.1 cm]D.west) -- ([yshift=-0.1 cm]C.east) ;

        \draw[->,dashed] ([xshift=0.1 cm]A.south) -- ([xshift=0.1 cm]C.north) ;
        \draw[->,dashed] ([xshift=-0.1 cm]C.north) -- ([xshift=-0.1 cm]A.south) ;

        \draw[->,dashed] ([xshift=0.1 cm]B.south) -- ([xshift=0.1 cm]D.north) ;
        \draw[->,dashed] ([xshift=-0.1 cm]D.north) -- ([xshift=-0.1 cm]B.south) ;

        \end{tikzpicture}
    }
    \subfigure[HCP]{
        \label{fig:edge.usage.hcp}
        \begin{tikzpicture}[node distance=15mm, thick]
        \tikzstyle{every node}=[minimum size=7mm]

        \node [fill=black!30!white ,draw=black,rectangle] (A) {5};
        \node [fill=white,draw=black,circle] (B) [right of=A] {4.3};
        \node [fill=black!30!white ,draw=black,rectangle] (C) [below of=A] {3};
        \node [fill=white,draw=black,circle] (D) [below of=B] {3.7};

        \draw[->] ([yshift=0.1 cm]A.east) -- ([yshift=0.1 cm]B.west) ;
        \draw[->,dashed] ([yshift=-0.1 cm]B.west) -- ([yshift=-0.1 cm]A.east) ;

        \draw[->] ([yshift=0.1 cm]C.east) -- ([yshift=0.1 cm]D.west) ;
        \draw[->,dashed] ([yshift=-0.1 cm]D.west) -- ([yshift=-0.1 cm]C.east) ;

        \draw[->,dashed] ([xshift=0.1 cm]A.south) -- ([xshift=0.1 cm]C.north) ;
        \draw[->,dashed] ([xshift=-0.1 cm]C.north) -- ([xshift=-0.1 cm]A.south) ;

        \draw[->] ([xshift=0.1 cm]B.south) -- ([xshift=0.1 cm]D.north) ;
        \draw[->] ([xshift=-0.1 cm]D.north) -- ([xshift=-0.1 cm]B.south) ;

        \end{tikzpicture}
    }
    \subfigure[SFR]{
        \label{fig:edge.usage.ihcp}
        \begin{tikzpicture}[node distance=15mm, thick]
        \tikzstyle{every node}=[minimum size=7mm]

        \node [fill=black!30!white ,draw=black,rectangle] (A) {5};
        \node [fill=white,draw=black,circle] (B) [right of=A] {3};
        \node [fill=black!30!white ,draw=black,circle] (C) [below of=A] {3};
        \node [fill=white,draw=black,rectangle] (D) [below of=B] {1};

        \draw[->] ([yshift=0.1 cm]A.east) -- ([yshift=0.1 cm]B.west) ;
        \draw[->,dashed] ([yshift=-0.1 cm]B.west) -- ([yshift=-0.1 cm]A.east) ;

        \draw[->,dashed] ([yshift=0.1 cm]C.east) -- ([yshift=0.1 cm]D.west) ;
        \draw[->] ([yshift=-0.1 cm]D.west) -- ([yshift=-0.1 cm]C.east) ;

        \draw[->] ([xshift=0.1 cm]A.south) -- ([xshift=0.1 cm]C.north) ;
        \draw[->,dashed] ([xshift=-0.1 cm]C.north) -- ([xshift=-0.1 cm]A.south) ;

        \draw[->,dashed] ([xshift=0.1 cm]B.south) -- ([xshift=0.1 cm]D.north) ;
        \draw[->] ([xshift=-0.1 cm]D.north) -- ([xshift=-0.1 cm]B.south) ;

        \end{tikzpicture}
    }
\caption{Different leverage of the linearity assumption by three approaches. $4$ items make an item-item network. The left two items are observed to be rated $5$-star and $3$-star respectively, as
drawn in grey. The right two items wait predicting, as drawn in
white. Different approaches consider those items as sources (drawn
in boxes) or non-sources (drawn in circles), where a non-source item
is expected to satisfy the linearity assumption. A
solid directed edge indicates that the rating on the tail is collected to calculate the second derivative of the rating on the head, while a dash arrow indicates the rating on the tail is not involved in the calculation on the head. (a) The kNN collaborative filtering calculates the
second derivative on unrated items only, and the calculation is
based on observed neighbors. (b) HCP calculates the second
derivative on unrated items only, and the calculation is based on
all neighbors. (c) SFR calculates the second derivative on rated
and unrated items except a few sources, and the calculation is based
on all neighbors. SFR makes use of much more neighborhood relations
than kNN and HCP.}
\label{fig:edge.usage}
\end{figure}
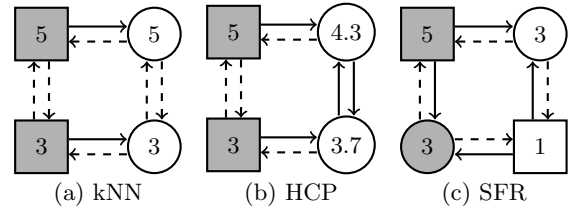

As discussed earlier, the kNN collaborative filtering technique (denoted as kNN in the rest of the paper) leverages the linearity assumption to estimate an unobserved rating. However, it actually calculates the second derivative with observed ratings of neighbors only, instead of the observed ratings and estimated rating of all the neighbors, as

\begin{equation}
\hat{r}_u(i) = \frac{\sum_{j\in N(i)\cap A(u)} w(i,j) r_u(j)}{\sum_{j\in N(i)\cap A(u)} w(i,j)},
\end{equation}
where $A(u)$ denotes the set of items that user $u$ has rated. The above calculation is an approximation, and could be unreliable when the observations are sparse, which is quite common in a typical recommender system. An example is shown in Figure \ref{fig:edge.usage} to demonstrate. In the example network consisting of $4$ items, the left two items in grey are observed to be rated $5$-star and $3$-star respectively, while the right two ones in white wait predicting. As shown in Figure \ref{fig:edge.usage.knn}, kNN collaborative filtering considers Equation (\ref{equ:2nd-derivative-vanish}) holding on unrated items as drawn in circles. Each unobserved rating is estimated with observed neighbors. The upper right item has only one neighbor observed, the upper left item, and therefore its estimation simply equals $5$-star rating. The estimation is bounded by the $5$-star rating on its observed neighbor.

The heat conduction process \cite{heat-conduction} (denoted as HCP in the rest of the paper) points out the shortcoming of kNN approach and improves the kNN approach by calculating the second derivative with all neighbor ratings, no matter observed or unobserved. In order to break the dilemma that a pair of unobserved neighboring items wait for each other to complete the calculation first, the approach simultaneously estimate all unobserved ratings by solving a linear system consisting of Equation (\ref{equ:2nd-derivative-vanish}) on all unobserved items. However, the requirement that Equation (\ref{equ:2nd-derivative-vanish}) holds on all unobserved items is a strong assumption and therefore limits its performance.

As shown in Figure \ref{fig:edge.usage.hcp}, HCP also considers Equation (\ref{equ:2nd-derivative-vanish}) holding on unrated items as drawn in circles. Different from kNN, an unobserved rating is estimated with observed and unobserved neighbors. The predicted rating of upper right item is calculated with its two neighbors, the observed upper left item and the unobserved lower right item. Solving Equation (\ref{equ:2nd-derivative-vanish}) on the right two items simultaneously results in $4.3$-star and $3.7$-star ratings on them respectively. Both estimators are bounded in the range of $[3,5]$.

In our view of scalar function recovery, in order to estimate the second derivative as accurately as possible, the second derivative is also calculated using all neighbor ratings, no matter observed or unobserved. Similarly, the recovery process also solves a bunch of instances of Equation (\ref{equ:2nd-derivative-vanish}) simultaneously to avoid the dilemma of mutually waiting. Different from HCP, we expect Equation (\ref{equ:2nd-derivative-vanish}) hold on most items, no matter observed or unobserved, instead of all unobserved items. The distinct benefit of our method is explained as follows.

Since an item with non-zero value of second derivative is probably rated higher than neighbor items (local maxima of a rating function) or lower than neighbor items (local minima), it might represent the point a user loves or hates the most in a local range of dozens of items. The rating on such an item could be unobserved due to many reasons. For example, the user is lazy to label her favorite movie on a website since she registers for a dozen movie websites, or a user is reluctant to post a rating to an extremely boring book, \mark{or the rating is actually observed but hidden in the testing set, }etc. Allowing Equation (\ref{equ:2nd-derivative-vanish}) not to hold on an unobserved item, we keep the possibility to view the unobserved item as a local favorite or dislike. Symmetrically, since an unobserved item might be a local maximum or minimum, it is also possible that an observed item is not a local maximum or minimum, and furthermore it has a chance to have a zero valued second derivative. Yet we do not exclude the possibility that Equation (\ref{equ:2nd-derivative-vanish}) holds on an observed item.

As shown in Figure \ref{fig:edge.usage.ihcp}, SFR considers Equation (\ref{equ:2nd-derivative-vanish}) holding on some items. For example we take the upper left and lower right items as sources, drawn in boxes. The remaining two items, drawn in circles, are not considered as sources and are expected to satisfy Equation (\ref{equ:2nd-derivative-vanish}). The missing ratings are predicted by simultaneously solving Equation (\ref{equ:2nd-derivative-vanish}) on the two non-source items, whose second derivatives are calculated with both observed and unobserved neighbors, as the arrows indicate. The results are $3$-star and $1$-star respectively, not bounded in the range of observed ratings.

In the demo all approaches predict with two instances of Equation (\ref{equ:2nd-derivative-vanish}). In actual, since our approach seeks for a solution minimizing the number of sources (discussed later), it is expected that the number of sources might be even smaller than the number of observed ones, and therefore the equations our approach uses might be more than the equations kNN and HCP make use of. More equations could provide more evidence to accurately recover a scalar function.

\begin{table}[h]\centering
\caption{Different leverage of the linearity assumption by three approaches.}
\begin{tabular}[c]{|c|c|c|}
  \hline
     & Calculates with & Calculates for \\
  \hline
    kNN & Observed neighbors & All unobserved ratings\\
  \hline
    HCP & All neighbors & All unobserved ratings\\
  \hline
    SFR & All neighbors & \multirow{2}{1.5in}{Most items (observed or unobserved)}\\
    & & \\
  \hline
\end{tabular}
\end{table}

\subsection{Connection to the rating bound problem}

Let us analyze the three views, kNN, HCP and our scalar function recovery, from the perspective of the rating bound problem.

\textbf{kNN: locally bounded.} A kNN prediction of an unobserved rating is bounded by the ratings of observed neighbor items, since kNN estimates an unobserved rating with the weighted average of ratings among observed neighbors. For example, if an unobserved rating has all neighbor items observed with $1$-star or $2$-star ratings by the same user, it has no chance to be predicted with a $3$-star rating. Therefore kNN can never correctly predict the rating on that item if it happens to be the user's most favorite movie.

\textbf{HCP: globally bounded.} An HCP prediction of an unobserved rating is not bounded by neighbor items, since it calculates the weighted average among all neighbors no matter observed or unobserved. However, in the HCP calculation it is required that the second derivatives vanish on all unobserved items. Therefore only observed items can have non-zero values of second derivative. As discussed earlier, local minima and maxima must have non-zero values of second derivative. As a result, all predictions have a global bound, i.e., they are bounded within the range between the lowest rating and the highest rating among observed ratings. Take the same example, even though an unobserved rating has all neighbor items observed with $1$-star or $2$-star ratings by the same user, it still has a chance to be predicted with a $3$-star rating as long as she has ever posted a $4$-star rating on any item. The HCP approach improves kNN by expanding the bounds from local range to global range. However, it still has no chance to receive a $5$-star prediction if the user has not posted a rating higher than $4$-star.

\textbf{SFR: not bounded.} In our approach, prediction on an unobserved rating is not bounded. Since the linearity assumption does not require all unobserved ratings satisfy Equation (\ref{equ:2nd-derivative-vanish}), a prediction might be higher/lower than all neighboring and far-away observations. Besides, we also allow an observed rating satisfying Equation (\ref{equ:2nd-derivative-vanish}) sometimes. Even the highest observed rating also has a chance to satisfy (\ref{equ:2nd-derivative-vanish}), i.e., at least one prediction among its neighbors is assigned a higher rating than it and thus breaks the rating bound. Therefore our approach does not suffer from the rating bound problem and thus has the potential to gain better performance on prediction, especially on items whose ground truth is indeed higher/lower than all observed ratings.

\section{Inference}\label{section:inference}

In this section we introduce our approach to recover a scalar function with the property of its second derivative.

The scalar function recovery process is formulated as the following inference job. Given a user $u$, we observe the value of her scalar function $r_u(\cdot)$ on a set of items $A(u)$ that she has posted ratings, and are required to estimate the value of $r_u(\cdot)$ on other items. We denote $\tilde{r}_u(i)$ for an observed rating that user $u$ posts on item $i$, and $\hat{r}_u(i)$ for an estimated rating. To complete the job, we search for a feasible solution with a minimal number of sources as follows,

\begin{equation}\label{equ:inference-with-l0}
\begin{split}
\hat{R}_u & = \arg\min_{R_u} ||\nabla^2 R_u||_0 \\
s.t. & R_{u,i} = \tilde{r}_u(i), \forall i \in O(u) \\
& R_{u,i} \in [c_l, c_h], \forall u,i
\end{split}
\end{equation}
where $\bm{R_u}$ is again a vector consisting of $r_u(\cdot)$ values. Since our approach does not bound prediction range with observations, one needs to specify the legal range $[c_l, c_h]$ a rating is allowed to be. A typical recommender system like MovieLens and Netflix requires a legal rating be between $[1,5]$, and in Yahoo! music the range is $[0, 100]$. The $l_0$-norm of $\nabla^2 R_u$, which is equivalent with $\nabla^2 r_u$, counts the number of sources. The solution comes by minimizing the number of sources under the hard constraint of observations within the predefined boundary. Unfortunately, the $l_0$-norm of a vector is difficult to minimize since it has no explicit form of gradient. For computational ease, we replace problem (\ref{equ:inference-with-l0}) with a slightly different form as follows

\begin{equation}\label{equ:inference}
\begin{split}
\hat{R}_u & = \arg\min_{R_u} ||\nabla^2 R_u||_p \\
s.t. & R_{u,i} = \tilde{r}_u(i), \forall i \in O(u) \\
& R_{u,i} \in [c_l, c_h], \forall u,i
\end{split}
\end{equation}
where $0<p<1$ is a parameter and $||\cdot||_p$ denotes $l_p$-norm of a vector $\vec{w}$ as follows

$$
||\vec{w}||_p = (\sum_k |w_k|^p)^\frac{1}{p},
$$
where $w_k$ is the $k^{th}$ element in $\vec{w}$. Since it is convenient to calculate the gradient of the $l_p$-norm of a vector, we can easily apply standard optimization algorithms such as gradient descent to find the solution. The gradient of objective function is

$$
\nabla_i ||\nabla^2 R_u||_p = (\sum_k |{(\nabla^2 R_u)}_k|^p)^{\frac{1}{p}-1} L_i^T v,
$$
where $\nabla_i$ indicates we only calculate gradient for unobserved ratings $R_{u,i},i\notin O(u)$, $(\nabla^2 R_u)_k$ is the $k^{th}$ element in the vector $\nabla^2 R_u$, $L_i$ consists of column vectors in the Laplacian matrix corresponding to unobserved items, and $L_i^T$ is its transpose. $v$ is a vector of the same size as $\nabla^2 R_u$ and each of its element $v_k=|{(\nabla^2 R_u)}_k|^{p-1} sign({(\nabla^2 R_u)}_k)$. In practice we set $p=\frac{1}{2}$ in later experiments. In Section \ref{section:discuss.p}, we will discuss the selection of p.

\subsection{Sources of a scalar function}

Why do we solve the scalar function recovery task by minimizing the number of sources? \mark{To explain, we first consider another question. As is discussed in Section \ref{subsection:leverage-the-assumption}, in our approach any item has a chance to satisfy Equation (\ref{equ:2nd-derivative-vanish}). Then on what items the equation holds? Since collaborative filtering applies Equation (\ref{equ:resnick}) to estimate ratings on all tasks and gains success in the last decade, it is reasonable to consider the above equation hold on all items. Unfortunately, a}A scalar function on a finite-sized network with all-zero second derivative must be a constant function, which is obviously not practical in a real recommender system. Therefore for any user, there must exist some items with non-zero values of the second derivative of her scalar function. We claim there should only exist as few sources as possible, and the function can be recovered by minimizing the number of those items. This claim is reasonable for the following two reasons.

Intuitively, every user has several favorite movies or books that she rates higher than neighbor items (local maxima of a rating function). Symmetrically she might also have several dislikes that she rates lower than neighbor items (local minima). Since her ratings on those items have no chance to be equal to the weighted average among neighbor items, the second derivative cannot vanish on those items. Therefore, the sources provide a superset of the items she loves or hates the most. By minimizing the number of sources, we lower the number of favorites and dislikes in an estimated interest distribution, capturing the intuition that it is quite unlikely a user has thousands of ``favorite'' movies.

Mathematically, if we view the scalar function as a scalar field defined on an item-item network, the second derivative is the divergence field of its gradient field, and the sources are points where the scalar function changes its gradient. The more frequently a scalar function changes its gradient, the more complex it could be. The number of sources could thus be considered a metric of function complexity or structural risk. Given partial observation of a scalar function, there might exist infinitely many feasible solutions that fit the observations. Thus minimizing the number of sources helps us to find a solution with the minimal risk of over-fitting among all feasible ones.



\subsection{Case study on toy data}

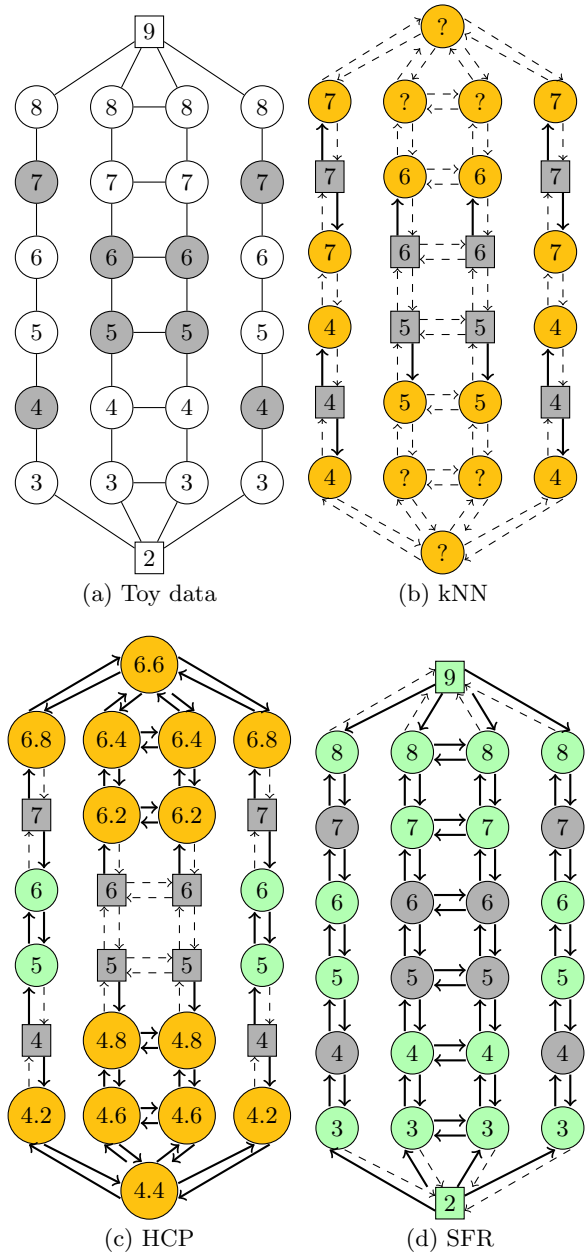
\begin{figure}
    \centering
    \subfigure[Toy data]{
        \label{fig:toydata.ground.truth}
        \begin{tikzpicture}[node distance=10mm]
        \tikzstyle{every node}=[fill=white,draw=black,minimum size=2mm,circle]

        \node (A1) [rectangle]   {2};
        \node (A3) [above of=A1, xshift= -5mm] {3};
        \node (A2) [left  of=A3] {3};
        \node (A4) [above of=A1, xshift= 5mm] {3};
        \node (A5) [right of=A4] {3};
        \node (A6) [above of=A2, fill=black!30!white] {4};
        \node (A7) [above of=A3] {4};
        \node (A8) [above of=A4] {4};
        \node (A9) [above of=A5, fill=black!30!white] {4};
        \node (A10) [above of=A6] {5};
        \node (A11) [above of=A7, fill=black!30!white] {5};
        \node (A12) [above of=A8, fill=black!30!white] {5};
        \node (A13) [above of=A9] {5};
        \node (A14) [above of=A10] {6};
        \node (A15) [above of=A11, fill=black!30!white] {6};
        \node (A16) [above of=A12, fill=black!30!white] {6};
        \node (A17) [above of=A13] {6};
        \node (A18) [above of=A14, fill=black!30!white] {7};
        \node (A19) [above of=A15] {7};
        \node (A20) [above of=A16] {7};
        \node (A21) [above of=A17, fill=black!30!white] {7};
        \node (A22) [above of=A18] {8};
        \node (A23) [above of=A19] {8};
        \node (A24) [above of=A20] {8};
        \node (A25) [above of=A21] {8};
        \node (A26) [above of=A23, xshift= 5mm, rectangle] {9};

        \draw[-] (A1)--(A2);
        \draw[-] (A1)--(A3);
        \draw[-] (A1)--(A4);
        \draw[-] (A1)--(A5);
        \draw[-] (A2)--(A6);
        \draw[-] (A3)--(A7);
        \draw[-] (A4)--(A8);
        \draw[-] (A5)--(A9);
        \draw[-] (A6)--(A10);
        \draw[-] (A7)--(A11);
        \draw[-] (A8)--(A12);
        \draw[-] (A9)--(A13);
        \draw[-] (A10)--(A14);
        \draw[-] (A11)--(A15);
        \draw[-] (A12)--(A16);
        \draw[-] (A13)--(A17);
        \draw[-] (A14)--(A18);
        \draw[-] (A15)--(A19);
        \draw[-] (A16)--(A20);
        \draw[-] (A17)--(A21);
        \draw[-] (A18)--(A22);
        \draw[-] (A19)--(A23);
        \draw[-] (A20)--(A24);
        \draw[-] (A21)--(A25);
        \draw[-] (A22)--(A26);
        \draw[-] (A23)--(A26);
        \draw[-] (A24)--(A26);
        \draw[-] (A25)--(A26);
        \draw[-] (A3)--(A4);
        \draw[-] (A7)--(A8);
        \draw[-] (A11)--(A12);
        \draw[-] (A15)--(A16);
        \draw[-] (A19)--(A20);
        \draw[-] (A23)--(A24);
        \end{tikzpicture}
    }
    \subfigure[kNN]{
        \label{fig:toydata.knn}
        \begin{tikzpicture}[node distance=10mm]
        \tikzstyle{every node}=[fill=yellow!50!orange,draw=black,minimum size=2mm,circle]

        \node (A1) {?};
        \node (A3) [above of=A1, xshift= -5mm] {?};
        \node (A2) [left  of=A3] {4};
        \node (A4) [above of=A1, xshift= 5mm] {?};
        \node (A5) [right of=A4] {4};
        \node (A6) [above of=A2, fill=black!30!white, rectangle] {4};
        \node (A7) [above of=A3] {5};
        \node (A8) [above of=A4] {5};
        \node (A9) [above of=A5, fill=black!30!white, rectangle] {4};
        \node (A10) [above of=A6] {4};
        \node (A11) [above of=A7, fill=black!30!white, rectangle] {5};
        \node (A12) [above of=A8, fill=black!30!white, rectangle] {5};
        \node (A13) [above of=A9] {4};
        \node (A14) [above of=A10] {7};
        \node (A15) [above of=A11, fill=black!30!white, rectangle] {6};
        \node (A16) [above of=A12, fill=black!30!white, rectangle] {6};
        \node (A17) [above of=A13] {7};
        \node (A18) [above of=A14, fill=black!30!white, rectangle] {7};
        \node (A19) [above of=A15] {6};
        \node (A20) [above of=A16] {6};
        \node (A21) [above of=A17, fill=black!30!white, rectangle] {7};
        \node (A22) [above of=A18] {7};
        \node (A23) [above of=A19] {?};
        \node (A24) [above of=A20] {?};
        \node (A25) [above of=A21] {7};
        \node (A26) [above of=A23, xshift= 5mm] {?};

        \draw[->,dashed] ([yshift=-1mm]A1.west)--([xshift=-1mm]A2.south);
        \draw[->,dashed] ([xshift=1mm]A2.south)--([yshift=1mm]A1.west);
        \draw[->,dashed] ([xshift=-3mm]A1.north)--([xshift=-1mm]A3.south);
        \draw[->,dashed] ([xshift= 1mm]A3.south)--([xshift=-1mm]A1.north);
        \draw[->,dashed] ([xshift= 1mm]A1.north)--([xshift=-1mm]A4.south);
        \draw[->,dashed] ([xshift= 1mm]A4.south)--([xshift= 3mm]A1.north);
        \draw[->,dashed] ([yshift= 1mm]A1.east)--([xshift=-1mm]A5.south);
        \draw[->,dashed] ([xshift= 1mm]A5.south)--([yshift=-1mm]A1.east);

        \draw[->,dashed] ([xshift=-1mm]A2.north)--([xshift=-1mm]A6.south);
        \draw[->,thick ] ([xshift= 1mm]A6.south)--([xshift= 1mm]A2.north);
        \draw[->,dashed] ([xshift=-1mm]A3.north)--([xshift=-1mm]A7.south);
        \draw[->,dashed] ([xshift= 1mm]A7.south)--([xshift= 1mm]A3.north);
        \draw[->,dashed] ([xshift=-1mm]A4.north)--([xshift=-1mm]A8.south);
        \draw[->,dashed] ([xshift= 1mm]A8.south)--([xshift= 1mm]A4.north);
        \draw[->,dashed] ([xshift=-1mm]A5.north)--([xshift=-1mm]A9.south);
        \draw[->,thick ] ([xshift= 1mm]A9.south)--([xshift= 1mm]A5.north);
        \draw[->,thick ] ([xshift=-1mm]A6.north)--([xshift=-1mm]A10.south);
        \draw[->,dashed] ([xshift= 1mm]A10.south)--([xshift= 1mm]A6.north);
        \draw[->,dashed] ([xshift=-1mm]A7.north)--([xshift=-1mm]A11.south);
        \draw[->,thick ] ([xshift= 1mm]A11.south)--([xshift= 1mm]A7.north);
        \draw[->,dashed] ([xshift=-1mm]A8.north)--([xshift=-1mm]A12.south);
        \draw[->,thick ] ([xshift= 1mm]A12.south)--([xshift= 1mm]A8.north);
        \draw[->,thick ] ([xshift=-1mm]A9.north)--([xshift=-1mm]A13.south);
        \draw[->,dashed] ([xshift= 1mm]A13.south)--([xshift= 1mm]A9.north);
        \draw[->,dashed] ([xshift=-1mm]A10.north)--([xshift=-1mm]A14.south);
        \draw[->,dashed] ([xshift= 1mm]A14.south)--([xshift= 1mm]A10.north);
        \draw[->,dashed] ([xshift=-1mm]A11.north)--([xshift=-1mm]A15.south);
        \draw[->,dashed] ([xshift= 1mm]A15.south)--([xshift= 1mm]A11.north);
        \draw[->,dashed] ([xshift=-1mm]A12.north)--([xshift=-1mm]A16.south);
        \draw[->,dashed] ([xshift= 1mm]A16.south)--([xshift= 1mm]A12.north);
        \draw[->,dashed] ([xshift=-1mm]A13.north)--([xshift=-1mm]A17.south);
        \draw[->,dashed] ([xshift= 1mm]A17.south)--([xshift= 1mm]A13.north);
        \draw[->,dashed] ([xshift=-1mm]A14.north)--([xshift=-1mm]A18.south);
        \draw[->,thick ] ([xshift= 1mm]A18.south)--([xshift= 1mm]A14.north);
        \draw[->,thick ] ([xshift=-1mm]A15.north)--([xshift=-1mm]A19.south);
        \draw[->,dashed] ([xshift= 1mm]A19.south)--([xshift= 1mm]A15.north);
        \draw[->,thick ] ([xshift=-1mm]A16.north)--([xshift=-1mm]A20.south);
        \draw[->,dashed] ([xshift= 1mm]A20.south)--([xshift= 1mm]A16.north);
        \draw[->,dashed] ([xshift=-1mm]A17.north)--([xshift=-1mm]A21.south);
        \draw[->,thick ] ([xshift= 1mm]A21.south)--([xshift= 1mm]A17.north);
        \draw[->,thick ] ([xshift=-1mm]A18.north)--([xshift=-1mm]A22.south);
        \draw[->,dashed] ([xshift= 1mm]A22.south)--([xshift= 1mm]A18.north);
        \draw[->,dashed] ([xshift=-1mm]A19.north)--([xshift=-1mm]A23.south);
        \draw[->,dashed] ([xshift= 1mm]A23.south)--([xshift= 1mm]A19.north);
        \draw[->,dashed] ([xshift=-1mm]A20.north)--([xshift=-1mm]A24.south);
        \draw[->,dashed] ([xshift= 1mm]A24.south)--([xshift= 1mm]A20.north);
        \draw[->,thick ] ([xshift=-1mm]A21.north)--([xshift=-1mm]A25.south);
        \draw[->,dashed] ([xshift= 1mm]A25.south)--([xshift= 1mm]A21.north);

        \draw[->,dashed] ([yshift=-1mm]A26.west) --([xshift= 1mm]A22.north);
        \draw[->,dashed] ([xshift=-1mm]A22.north)--([yshift= 1mm]A26.west) ;
        \draw[->,dashed] ([xshift=-1mm]A23.north)--([xshift=-3mm]A26.south);
        \draw[->,dashed] ([xshift=-1mm]A26.south)--([xshift= 1mm]A23.north);
        \draw[->,dashed] ([xshift=-1mm]A24.north)--([xshift= 1mm]A26.south);
        \draw[->,dashed] ([xshift= 3mm]A26.south)--([xshift= 1mm]A24.north);
        \draw[->,dashed] ([yshift= 1mm]A26.east) --([xshift= 1mm]A25.north);
        \draw[->,dashed] ([xshift=-1mm]A25.north)--([yshift=-1mm]A26.east) ;

        \draw[->,dashed] ([yshift= 1mm]A3.east)--([yshift= 1mm]A4.west);
        \draw[->,dashed] ([yshift=-1mm]A4.west)--([yshift=-1mm]A3.east);
        \draw[->,dashed] ([yshift= 1mm]A7.east)--([yshift= 1mm]A8.west);
        \draw[->,dashed] ([yshift=-1mm]A8.west)--([yshift=-1mm]A7.east);
        \draw[->,dashed] ([yshift= 1mm]A11.east)--([yshift= 1mm]A12.west);
        \draw[->,dashed] ([yshift=-1mm]A12.west)--([yshift=-1mm]A11.east);
        \draw[->,dashed] ([yshift= 1mm]A15.east)--([yshift= 1mm]A16.west);
        \draw[->,dashed] ([yshift=-1mm]A16.west)--([yshift=-1mm]A15.east);
        \draw[->,dashed] ([yshift= 1mm]A19.east)--([yshift= 1mm]A20.west);
        \draw[->,dashed] ([yshift=-1mm]A20.west)--([yshift=-1mm]A19.east);
        \draw[->,dashed] ([yshift= 1mm]A23.east)--([yshift= 1mm]A24.west);
        \draw[->,dashed] ([yshift=-1mm]A24.west)--([yshift=-1mm]A23.east);

        \end{tikzpicture}
    }

    \subfigure[HCP]{
        \label{fig:toydata.hcp}
        \begin{tikzpicture}[node distance=10mm]
        \tikzstyle{every node}=[fill=yellow!50!orange,draw=black,minimum size=2mm,circle]

        \node (A1) {4.4};
        \node (A3) [above of=A1, xshift= -5mm] {4.6};
        \node (A4) [above of=A1, xshift=  5mm] {4.6};
        \node (A2) [left  of=A3] {4.2};
        \node (A5) [right of=A4] {4.2};
        \node (A6) [above of=A2, fill=black!30!white, rectangle] {4};
        \node (A7) [above of=A3] {4.8};
        \node (A8) [above of=A4] {4.8};
        \node (A9) [above of=A5, fill=black!30!white, rectangle] {4};
        \node (A10) [above of=A6, fill=green!30!white] {5};
        \node (A11) [above of=A7, fill=black!30!white, rectangle] {5};
        \node (A12) [above of=A8, fill=black!30!white, rectangle] {5};
        \node (A13) [above of=A9, fill=green!30!white] {5};
        \node (A14) [above of=A10, fill=green!30!white] {6};
        \node (A15) [above of=A11, fill=black!30!white, rectangle] {6};
        \node (A16) [above of=A12, fill=black!30!white, rectangle] {6};
        \node (A17) [above of=A13, fill=green!30!white] {6};
        \node (A18) [above of=A14, fill=black!30!white, rectangle] {7};
        \node (A19) [above of=A15] {6.2};
        \node (A20) [above of=A16] {6.2};
        \node (A21) [above of=A17, fill=black!30!white, rectangle] {7};
        \node (A22) [above of=A18] {6.8};
        \node (A23) [above of=A19] {6.4};
        \node (A24) [above of=A20] {6.4};
        \node (A25) [above of=A21] {6.8};
        \node (A26) [above of=A23, xshift= 5mm] {6.6};

        \draw[->,thick ] ([yshift=-1mm]A1.west) --([xshift=-1mm]A2.south);
        \draw[->,thick ] ([xshift= 1mm]A2.south)--([yshift= 1mm]A1.west);
        \draw[->,thick ] ([xshift=-3mm]A1.north)--([xshift=-1mm]A3.south);
        \draw[->,thick ] ([xshift= 1mm]A3.south)--([xshift=-1mm]A1.north);
        \draw[->,thick ] ([xshift= 1mm]A1.north)--([xshift=-1mm]A4.south);
        \draw[->,thick ] ([xshift= 1mm]A4.south)--([xshift= 3mm]A1.north);
        \draw[->,thick ] ([yshift= 1mm]A1.east) --([xshift=-1mm]A5.south);
        \draw[->,thick ] ([xshift= 1mm]A5.south)--([yshift=-1mm]A1.east);

        \draw[->,dashed] ([xshift=-1mm]A2.north)--([xshift=-1mm]A6.south);
        \draw[->,thick ] ([xshift= 1mm]A6.south)--([xshift= 1mm]A2.north);
        \draw[->,thick ] ([xshift=-1mm]A3.north)--([xshift=-1mm]A7.south);
        \draw[->,thick ] ([xshift= 1mm]A7.south)--([xshift= 1mm]A3.north);
        \draw[->,thick ] ([xshift=-1mm]A4.north)--([xshift=-1mm]A8.south);
        \draw[->,thick ] ([xshift= 1mm]A8.south)--([xshift= 1mm]A4.north);
        \draw[->,dashed] ([xshift=-1mm]A5.north)--([xshift=-1mm]A9.south);
        \draw[->,thick ] ([xshift= 1mm]A9.south)--([xshift= 1mm]A5.north);
        \draw[->,thick ] ([xshift=-1mm]A6.north)--([xshift=-1mm]A10.south);
        \draw[->,dashed] ([xshift= 1mm]A10.south)--([xshift= 1mm]A6.north);
        \draw[->,dashed] ([xshift=-1mm]A7.north)--([xshift=-1mm]A11.south);
        \draw[->,thick ] ([xshift= 1mm]A11.south)--([xshift= 1mm]A7.north);
        \draw[->,dashed] ([xshift=-1mm]A8.north)--([xshift=-1mm]A12.south);
        \draw[->,thick ] ([xshift= 1mm]A12.south)--([xshift= 1mm]A8.north);
        \draw[->,thick ] ([xshift=-1mm]A9.north)--([xshift=-1mm]A13.south);
        \draw[->,dashed] ([xshift= 1mm]A13.south)--([xshift= 1mm]A9.north);
        \draw[->,thick ] ([xshift=-1mm]A10.north)--([xshift=-1mm]A14.south);
        \draw[->,thick ] ([xshift= 1mm]A14.south)--([xshift= 1mm]A10.north);
        \draw[->,dashed] ([xshift=-1mm]A11.north)--([xshift=-1mm]A15.south);
        \draw[->,dashed] ([xshift= 1mm]A15.south)--([xshift= 1mm]A11.north);
        \draw[->,dashed] ([xshift=-1mm]A12.north)--([xshift=-1mm]A16.south);
        \draw[->,dashed] ([xshift= 1mm]A16.south)--([xshift= 1mm]A12.north);
        \draw[->,thick ] ([xshift=-1mm]A13.north)--([xshift=-1mm]A17.south);
        \draw[->,thick ] ([xshift= 1mm]A17.south)--([xshift= 1mm]A13.north);
        \draw[->,dashed] ([xshift=-1mm]A14.north)--([xshift=-1mm]A18.south);
        \draw[->,thick ] ([xshift= 1mm]A18.south)--([xshift= 1mm]A14.north);
        \draw[->,thick ] ([xshift=-1mm]A15.north)--([xshift=-1mm]A19.south);
        \draw[->,dashed] ([xshift= 1mm]A19.south)--([xshift= 1mm]A15.north);
        \draw[->,thick ] ([xshift=-1mm]A16.north)--([xshift=-1mm]A20.south);
        \draw[->,dashed] ([xshift= 1mm]A20.south)--([xshift= 1mm]A16.north);
        \draw[->,dashed] ([xshift=-1mm]A17.north)--([xshift=-1mm]A21.south);
        \draw[->,thick ] ([xshift= 1mm]A21.south)--([xshift= 1mm]A17.north);
        \draw[->,thick ] ([xshift=-1mm]A18.north)--([xshift=-1mm]A22.south);
        \draw[->,dashed] ([xshift= 1mm]A22.south)--([xshift= 1mm]A18.north);
        \draw[->,thick ] ([xshift=-1mm]A19.north)--([xshift=-1mm]A23.south);
        \draw[->,thick ] ([xshift= 1mm]A23.south)--([xshift= 1mm]A19.north);
        \draw[->,thick ] ([xshift=-1mm]A20.north)--([xshift=-1mm]A24.south);
        \draw[->,thick ] ([xshift= 1mm]A24.south)--([xshift= 1mm]A20.north);
        \draw[->,thick ] ([xshift=-1mm]A21.north)--([xshift=-1mm]A25.south);
        \draw[->,dashed] ([xshift= 1mm]A25.south)--([xshift= 1mm]A21.north);

        \draw[->,thick ] ([yshift=-1mm]A26.west) --([xshift= 1mm]A22.north);
        \draw[->,thick ] ([xshift=-1mm]A22.north)--([yshift= 1mm]A26.west) ;
        \draw[->,thick ] ([xshift=-1mm]A23.north)--([xshift=-3mm]A26.south);
        \draw[->,thick ] ([xshift=-1mm]A26.south)--([xshift= 1mm]A23.north);
        \draw[->,thick ] ([xshift=-1mm]A24.north)--([xshift= 1mm]A26.south);
        \draw[->,thick ] ([xshift= 3mm]A26.south)--([xshift= 1mm]A24.north);
        \draw[->,thick ] ([yshift= 1mm]A26.east) --([xshift= 1mm]A25.north);
        \draw[->,thick ] ([xshift=-1mm]A25.north)--([yshift=-1mm]A26.east) ;

        \draw[->,thick ] ([yshift= 1mm]A3.east)--([yshift= 1mm]A4.west);
        \draw[->,thick ] ([yshift=-1mm]A4.west)--([yshift=-1mm]A3.east);
        \draw[->,thick ] ([yshift= 1mm]A7.east)--([yshift= 1mm]A8.west);
        \draw[->,thick ] ([yshift=-1mm]A8.west)--([yshift=-1mm]A7.east);
        \draw[->,dashed] ([yshift= 1mm]A11.east)--([yshift= 1mm]A12.west);
        \draw[->,dashed] ([yshift=-1mm]A12.west)--([yshift=-1mm]A11.east);
        \draw[->,dashed] ([yshift= 1mm]A15.east)--([yshift= 1mm]A16.west);
        \draw[->,dashed] ([yshift=-1mm]A16.west)--([yshift=-1mm]A15.east);
        \draw[->,thick ] ([yshift= 1mm]A19.east)--([yshift= 1mm]A20.west);
        \draw[->,thick ] ([yshift=-1mm]A20.west)--([yshift=-1mm]A19.east);
        \draw[->,thick ] ([yshift= 1mm]A23.east)--([yshift= 1mm]A24.west);
        \draw[->,thick ] ([yshift=-1mm]A24.west)--([yshift=-1mm]A23.east);

        \end{tikzpicture}
    }
    \subfigure[SFR]{
        \label{fig:toydata.sfr}
       \begin{tikzpicture}[node distance=10mm]
        \tikzstyle{every node}=[fill=green!30!white,draw=black,minimum size=2mm,circle]

        \node (A1) [rectangle]   {2};
        \node (A3) [above of=A1, xshift= -5mm] {3};
        \node (A2) [left  of=A3] {3};
        \node (A4) [above of=A1, xshift= 5mm] {3};
        \node (A5) [right of=A4] {3};
        \node (A6) [above of=A2, fill=black!30!white] {4};
        \node (A7) [above of=A3] {4};
        \node (A8) [above of=A4] {4};
        \node (A9) [above of=A5, fill=black!30!white] {4};
        \node (A10) [above of=A6] {5};
        \node (A11) [above of=A7, fill=black!30!white] {5};
        \node (A12) [above of=A8, fill=black!30!white] {5};
        \node (A13) [above of=A9] {5};
        \node (A14) [above of=A10] {6};
        \node (A15) [above of=A11, fill=black!30!white] {6};
        \node (A16) [above of=A12, fill=black!30!white] {6};
        \node (A17) [above of=A13] {6};
        \node (A18) [above of=A14, fill=black!30!white] {7};
        \node (A19) [above of=A15] {7};
        \node (A20) [above of=A16] {7};
        \node (A21) [above of=A17, fill=black!30!white] {7};
        \node (A22) [above of=A18] {8};
        \node (A23) [above of=A19] {8};
        \node (A24) [above of=A20] {8};
        \node (A25) [above of=A21] {8};
        \node (A26) [above of=A23, xshift= 5mm, rectangle] {9};

        \draw[->,thick ] ([yshift=-1mm]A1.west) --([xshift=-1mm]A2.south);
        \draw[->,dashed] ([xshift= 1mm]A2.south)--([yshift= 1mm]A1.west);
        \draw[->,thick ] ([xshift=-3mm]A1.north)--([xshift=-1mm]A3.south);
        \draw[->,dashed] ([xshift= 1mm]A3.south)--([xshift=-1mm]A1.north);
        \draw[->,thick ] ([xshift= 1mm]A1.north)--([xshift=-1mm]A4.south);
        \draw[->,dashed] ([xshift= 1mm]A4.south)--([xshift= 3mm]A1.north);
        \draw[->,thick ] ([yshift= 1mm]A1.east) --([xshift=-1mm]A5.south);
        \draw[->,dashed] ([xshift= 1mm]A5.south)--([yshift=-1mm]A1.east);

        \draw[->,thick ] ([xshift=-1mm]A2.north)--([xshift=-1mm]A6.south);
        \draw[->,thick ] ([xshift= 1mm]A6.south)--([xshift= 1mm]A2.north);
        \draw[->,thick ] ([xshift=-1mm]A3.north)--([xshift=-1mm]A7.south);
        \draw[->,thick ] ([xshift= 1mm]A7.south)--([xshift= 1mm]A3.north);
        \draw[->,thick ] ([xshift=-1mm]A4.north)--([xshift=-1mm]A8.south);
        \draw[->,thick ] ([xshift= 1mm]A8.south)--([xshift= 1mm]A4.north);
        \draw[->,thick ] ([xshift=-1mm]A5.north)--([xshift=-1mm]A9.south);
        \draw[->,thick ] ([xshift= 1mm]A9.south)--([xshift= 1mm]A5.north);
        \draw[->,thick ] ([xshift=-1mm]A6.north)--([xshift=-1mm]A10.south);
        \draw[->,thick ] ([xshift= 1mm]A10.south)--([xshift= 1mm]A6.north);
        \draw[->,thick ] ([xshift=-1mm]A7.north)--([xshift=-1mm]A11.south);
        \draw[->,thick ] ([xshift= 1mm]A11.south)--([xshift= 1mm]A7.north);
        \draw[->,thick ] ([xshift=-1mm]A8.north)--([xshift=-1mm]A12.south);
        \draw[->,thick ] ([xshift= 1mm]A12.south)--([xshift= 1mm]A8.north);
        \draw[->,thick ] ([xshift=-1mm]A9.north)--([xshift=-1mm]A13.south);
        \draw[->,thick ] ([xshift= 1mm]A13.south)--([xshift= 1mm]A9.north);
        \draw[->,thick ] ([xshift=-1mm]A10.north)--([xshift=-1mm]A14.south);
        \draw[->,thick ] ([xshift= 1mm]A14.south)--([xshift= 1mm]A10.north);
        \draw[->,thick ] ([xshift=-1mm]A11.north)--([xshift=-1mm]A15.south);
        \draw[->,thick ] ([xshift= 1mm]A15.south)--([xshift= 1mm]A11.north);
        \draw[->,thick ] ([xshift=-1mm]A12.north)--([xshift=-1mm]A16.south);
        \draw[->,thick ] ([xshift= 1mm]A16.south)--([xshift= 1mm]A12.north);
        \draw[->,thick ] ([xshift=-1mm]A13.north)--([xshift=-1mm]A17.south);
        \draw[->,thick ] ([xshift= 1mm]A17.south)--([xshift= 1mm]A13.north);
        \draw[->,thick ] ([xshift=-1mm]A14.north)--([xshift=-1mm]A18.south);
        \draw[->,thick ] ([xshift= 1mm]A18.south)--([xshift= 1mm]A14.north);
        \draw[->,thick ] ([xshift=-1mm]A15.north)--([xshift=-1mm]A19.south);
        \draw[->,thick ] ([xshift= 1mm]A19.south)--([xshift= 1mm]A15.north);
        \draw[->,thick ] ([xshift=-1mm]A16.north)--([xshift=-1mm]A20.south);
        \draw[->,thick ] ([xshift= 1mm]A20.south)--([xshift= 1mm]A16.north);
        \draw[->,thick ] ([xshift=-1mm]A17.north)--([xshift=-1mm]A21.south);
        \draw[->,thick ] ([xshift= 1mm]A21.south)--([xshift= 1mm]A17.north);
        \draw[->,thick ] ([xshift=-1mm]A18.north)--([xshift=-1mm]A22.south);
        \draw[->,thick ] ([xshift= 1mm]A22.south)--([xshift= 1mm]A18.north);
        \draw[->,thick ] ([xshift=-1mm]A19.north)--([xshift=-1mm]A23.south);
        \draw[->,thick ] ([xshift= 1mm]A23.south)--([xshift= 1mm]A19.north);
        \draw[->,thick ] ([xshift=-1mm]A20.north)--([xshift=-1mm]A24.south);
        \draw[->,thick ] ([xshift= 1mm]A24.south)--([xshift= 1mm]A20.north);
        \draw[->,thick ] ([xshift=-1mm]A21.north)--([xshift=-1mm]A25.south);
        \draw[->,thick ] ([xshift= 1mm]A25.south)--([xshift= 1mm]A21.north);

        \draw[->,thick ] ([yshift=-1mm]A26.west) --([xshift= 1mm]A22.north);
        \draw[->,dashed] ([xshift=-1mm]A22.north)--([yshift= 1mm]A26.west) ;
        \draw[->,dashed] ([xshift=-1mm]A23.north)--([xshift=-3mm]A26.south);
        \draw[->,thick ] ([xshift=-1mm]A26.south)--([xshift= 1mm]A23.north);
        \draw[->,dashed] ([xshift=-1mm]A24.north)--([xshift= 1mm]A26.south);
        \draw[->,thick ] ([xshift= 3mm]A26.south)--([xshift= 1mm]A24.north);
        \draw[->,thick ] ([yshift= 1mm]A26.east) --([xshift= 1mm]A25.north);
        \draw[->,dashed] ([xshift=-1mm]A25.north)--([yshift=-1mm]A26.east) ;

        \draw[->,thick ] ([yshift= 1mm]A3.east)--([yshift= 1mm]A4.west);
        \draw[->,thick ] ([yshift=-1mm]A4.west)--([yshift=-1mm]A3.east);
        \draw[->,thick ] ([yshift= 1mm]A7.east)--([yshift= 1mm]A8.west);
        \draw[->,thick ] ([yshift=-1mm]A8.west)--([yshift=-1mm]A7.east);
        \draw[->,thick ] ([yshift= 1mm]A11.east)--([yshift= 1mm]A12.west);
        \draw[->,thick ] ([yshift=-1mm]A12.west)--([yshift=-1mm]A11.east);
        \draw[->,thick ] ([yshift= 1mm]A15.east)--([yshift= 1mm]A16.west);
        \draw[->,thick ] ([yshift=-1mm]A16.west)--([yshift=-1mm]A15.east);
        \draw[->,thick ] ([yshift= 1mm]A19.east)--([yshift= 1mm]A20.west);
        \draw[->,thick ] ([yshift=-1mm]A20.west)--([yshift=-1mm]A19.east);
        \draw[->,thick ] ([yshift= 1mm]A23.east)--([yshift= 1mm]A24.west);
        \draw[->,thick ] ([yshift=-1mm]A24.west)--([yshift=-1mm]A23.east);

        \end{tikzpicture}
    }
\caption{(Color online) Toy data demonstrating three approaches. Each node represents an item. An item shown in a box means that it is a source (or considered as a source), while an item shown in a circle means that it is not a source (or not considered as a source). A user's ratings are labeled on items. An observed rating is labeled in grey, while an unobserved rating is labeled in white. An accurately estimated rating is labeled in green, while an inaccurate estimation is labeled in orange. Each edge represents a relation between two items. An undirected edge in (a) indicates two items are related. A solid directed edge in (b)(c)(d) indicates that the edge is leveraged. Specifically, the rating on the tail is used to calculate the second derivative of the rating on the head. A dash directed edge indicates that the edge is not leveraged.}
\label{fig:toydata}
\end{figure}

We build an artificial dataset to demonstrate how our approach works. As shown in Figure \ref{fig:toydata.ground.truth}, the dataset contains a user's rating on $26$ items. The items, represented by circles and boxes, are connected with edges representing item relation (e.g., similarity between two items) with uniform weights. Ratings are labeled on items. The items she loves and dislikes the most locate at the top and bottom respectively, playing the role of sources (shown in boxes). Other items are assigned with proper ratings according to the linearity assumption (shown in circles). Ratings on $8$ items are observed (labeled in grey), and ratings on other items are waiting estimation (labeled in white). An accurately estimated rating will be labeled in green, while an inaccurate estimation will be labeled in orange.

We firstly run kNN and HCP on the toy dataset. kNN considers the observed ratings as sources and estimates each unobserved rating with the weighted average of ratings on its observed neighbors. As shown in Figure \ref{fig:toydata.knn}, kNN estimates the unobserved ratings with $12$ edges. Each unobserved rating has at most one neighbor observed, thus the average among observed neighbors provides an inaccurate estimator of the average among all neighbors. As a result, kNN inaccurately estimates all unobserved ratings. Besides, kNN fails to present estimations to unobserved items with no neighbors observed, labeled with question marks.

HCP also considers the observed ratings as sources and estimates unobserved ratings by requiring the linearity assumption satisfied on all unobserved items. As shown in Figure \ref{fig:toydata.hcp}, HCP estimates the unobserved ratings with $48$ edges. It outperforms kNN by successfully estimating unobserved ratings of the $4$ items in the middle, whose neighbors are either observed or easy to correctly estimate. However, due to the rating bound problem, HCP cannot accurately estimate the unobserved ratings of the $7$ items near the top, whose real ratings are higher than all observations. The similar problem occurs on the $7$ items near the bottom.


Our approach does not assume the observed ratings are sources. Instead, it seeks a solution to (approximately) minimize the number of unknown sources. The seeking converges to a solution that (correctly) takes the top and bottom items as two sources. It leverages $60$ edges to calculate the unobserved ratings, and accurately estimates all unobserved ratings as shown in Figure \ref{fig:toydata.sfr}. Although the observed ratings are narrowed in the range of $[4,7]$, our approach does not suffer from the rating bound problem and accurately recovers the ratings out of the range, especially ratings on the two sources.


\subsection{Connection with conventional approaches}

The conventional approaches such as kNN and HCP could be considered as special cases of our approach.

The kNN approach is a special case of our approach with two additional assumptions. First, the second derivative vanishes on all unobserved items. Compared with our linearity assumption that the second derivative vanishes on most items no matter observed or unobserved, the assumption of kNN is obviously stronger and therefore its prediction would be a subset of our feasible solution set. Second, calculation with partially observed neighbors provides a good estimator of the second derivative. In kNN, the weighted average among neighbors is not calculated with all neighbors, but instead with observed neighbors only. Such a way implies the belief that the estimator based on incomplete information provides an accurate approximation.

HCP extends the kNN by removing the second assumption. It requires the weighted average calculated with all neighbors, observed and unobserved, in order to achieve a more accurate estimation. However, it keeps the first assumption that requires the second derivative vanish on all unobserved items, which makes its feasible solution set also a subset of our feasible solution set.

To summarize, the two approaches kNN and HCP are built on assumptions stronger than ours, which makes them two special cases of our approach. Since their feasible solution sets are subsets of our feasible solution set, our approach has the potential to achieve better performance.

\subsection{The selection of parameter p}\label{section:discuss.p}

The selection of $p$ is an interesting topic when optimizing Equation (\ref{equ:inference}) to recover unobserved ratings. Different $p$ values might lead to different prediction results. The original Equation (\ref{equ:inference-with-l0}) corresponds to the case $p=0$, which is straightforward but difficult to solve explicitly. It provides an accurate solution if the network is small enough for an exhaustive search, and unfortunately fails to solve stably within an acceptable time period when the problem size becomes larger. $p \in (0,1)$ provides be a good approximation to the $p=0$ case but is much easier to optimize with an explicit form of gradient, and that is why we choose $p=\frac{1}{2}$ in our practice. When $p\geq 1$, the objective function does not lead to a sparse solution with very few sources. It is an open question that how the selection of $p$ influences the approximation.

\section{Experimental Validation}\label{section:experiments}

In this section we take experiments to study the rating bound problem and evaluate performance of different approaches on the problem.

\subsection{Data collection}\label{subection:dataset}
We empirically evaluate our approach on three datasets: Douban, Goodreads and MovieLens. Douban and Goodreads datasets are crawled from two online collection websites\footnote{www.douban.com and www.goodreads.com} where users rate millions of movies, books and music. After crawled the data, we removed inactive users since their inactivity may lead to unreliable statistics. The MovieLens dataset is a benchmark dataset in the latest decade. Statistics of those datasets are reported in Table \ref{table:data_collection}.

\begin{table}[ht]\centering
\caption{Datasets description}\label{table:data_collection}
\begin{tabular}[c]{|c|c|c|c|}
  \hline
     & Douban & Goodreads & MovieLens\\
  \hline
    Number of users & $32,384$ & $32,907$ & $6,040$\\
  \hline
    Number of items & $14,923$ & $13,548$ & $3,706$\\
  \hline
    Number of ratings & $345,293$ & $168,926$ & $1,000,209$\\
  \hline
\end{tabular}
\end{table}

For each dataset, $80\%$ examples (ratings) are taken as a training set, and the rest $20\%$ as a testing set. We calculate Pearson's correlation coefficient between ratings in the training set on each pair of items, and build an item-item network in which two items are linked with a weight equal to their correlation coefficient if the correlation is above $0.2$ (Douban and Goodreads) or $0.5$ (MovieLens).

\subsection{Studying the rating bound problem}
In order to demonstrate how the rating bound problem is critical to a recommender system, we count the examples suffering from the rating bound problem, i.e., where a rating is either higher or lower than all observed ratings on its neighbor items by the same user. Besides, we run kNN approach to predict the ratings in the testing set in order to examine to what extent those examples contribute to the prediction performance.

As reported in Table \ref{table:rating.bound.problem}, up to $15\%$ examples in the testing set of each dataset suffering from the rating bound problem. The real ratings on those examples are either higher or lower than all neighbor ratings in the training set, which means the kNN approach has no chance to correctly predict those ratings. Not surprisingly, those examples make up about $44\%$ of the prediction error among all testing examples\footnote{The prediction error is measured with squared residual rather than the standard evaluation metric RMSE (root-mean-square error), since it is not straightforward to declare the percentage that a part of examples contribute under a root calculation.}, indicating that it is much more difficult to correctly predict an example with rating bound problem than an example without the problem. Therefore the ability to solve the rating bound problem is critical to evaluate the performance of a recommender system.

\begin{table}[h]\centering
\caption{The number of examples suffering from rating bound problem in the testing set, as well as their contribution to the prediction error (measured in squared residual) calculated with a standard kNN approach.}
\label{table:rating.bound.problem}
\begin{tabular}{|c|c|c|c|c|}
\hline
& \multicolumn{2}{|c|}{\multirow{2}{1in}{Rating higher than neighbors}} & \multicolumn{2}{|c|}{\multirow{2}{1in}{Rating lower than neighbors}} \\
& \multicolumn{2}{c|}{} & \multicolumn{2}{c|}{} \\ \cline{2-5}
& \multirow{2}{*}{Count} & \multirow{2}{0.7in}{Error contribution}  & \multirow{2}{*}{Count} & \multirow{2}{0.7in}{Error contribution}\\
& & & & \\ \hline
Douban & 7.17\% & 18.28\% & 8.40\% & 25.57\% \\ \hline
Goodreads & 6.55\% & 17.66\% & 9.12\% & 27.02\% \\ \hline
MovieLens & 11.43\% &30.77\% &6.08\% & 14.12\% \\ \hline
\end{tabular}
\end{table}

\subsection{Empirical results}

We test our approach on real datasets to examine its ability to solve the rating bound problem, compared with kNN and HCP as baselines. Based on ratings in the training set and an item-item network, all approaches are tested to predict the ratings in the testing set which encounter the rating bound problem, i.e., a ground truth rating in the testing set is higher or lower than all neighbor ratings in the training set. The prediction results are evaluated with RMSE (root-mean-square error) and reported in Table \ref{table:result.rating.bound}, where ``Higher'' and ``Lower'' means the examples whose ratings are higher or lower than all observed neighbors respectively, and ``All'' means their combination. Our approach shows consistently better ability to solve the rating bound problem on different datasets, with a reduction on prediction error by up to $37\%$ compared with kNN. The improvement is even more significant on items whose ratings are higher than observed neighbors. Since an item rated higher than neighbors is probably a user's positive interest center, our approach is expected to achieve much better user experience by accurately discover a user's favorites.

\begin{table}[h!]\centering
\caption{Evaluating prediction error with RMSE on examples
encountering the rating bound problem. Our approach SFR consistently
achieves the lowest prediction error in different datasets and
different samples, with a reduction up to $37\%$ compared with kNN
collaborative filtering.}\label{table:result.rating.bound}
\begin{tabular}[c]{|c|c|c|c|}
  \hline
    \multicolumn{4}{|c|}{Douban} \\ \hline
    & kNN & HCP & SFR \\ \hline
    All & 1.532 & 1.287 & \textbf{1.269}  \\  \hline
    Higher & 1.458 & \textbf{1.086} & \textbf{1.086} \\  \hline
    Lower & 1.594 & 1.437 & \textbf{1.408} \\ \hline\hline
    \multicolumn{4}{|c|}{Goodreads} \\  \hline
    & kNN & HCP & SFR \\ \hline
    All & 1.566 & 1.404 & \textbf{1.355} \\ \hline
    Higher & 1.524 & 1.225 & \textbf{1.211} \\ \hline
    Lower & 1.597 & 1.521 & \textbf{1.450} \\ \hline\hline
    \multicolumn{4}{|c|}{MovieLens}\\  \hline
    & kNN & HCP & SFR \\ \hline
    All & 1.849 & 1.303 & \textbf{1.263} \\ \hline
    Higher & 1.895 & 1.204 & \textbf{1.185} \\ \hline
    Lower & 1.760 & 1.472 & \textbf{1.397} \\ \hline
\end{tabular}
\end{table}

\begin{figure*}[ht]
\centering
 \subfigure[Douban (higher)]{
    \includegraphics[scale = 0.4] {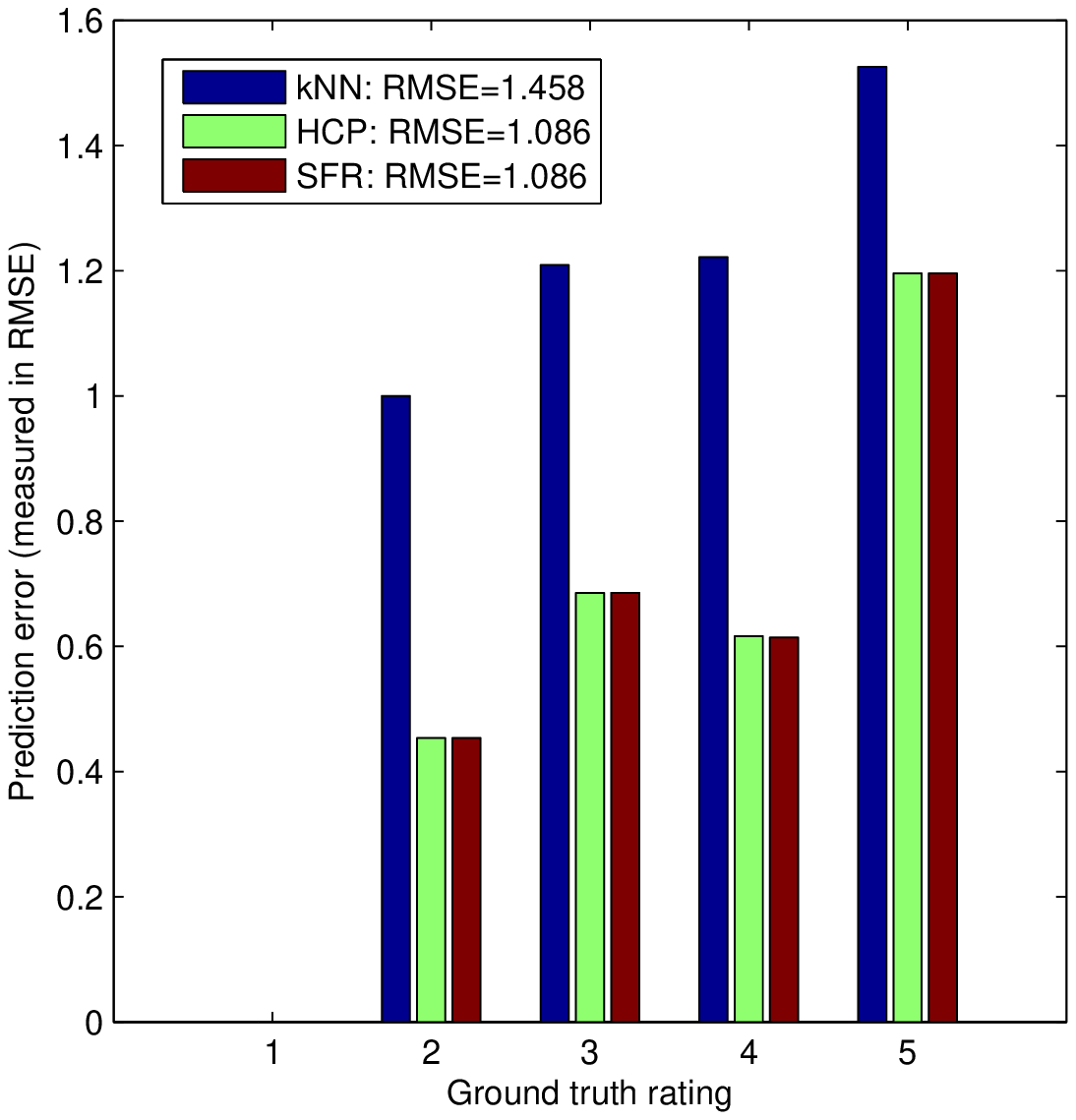}
 }
 \subfigure[Goodreads (higher)]{
    \includegraphics[scale = 0.4] {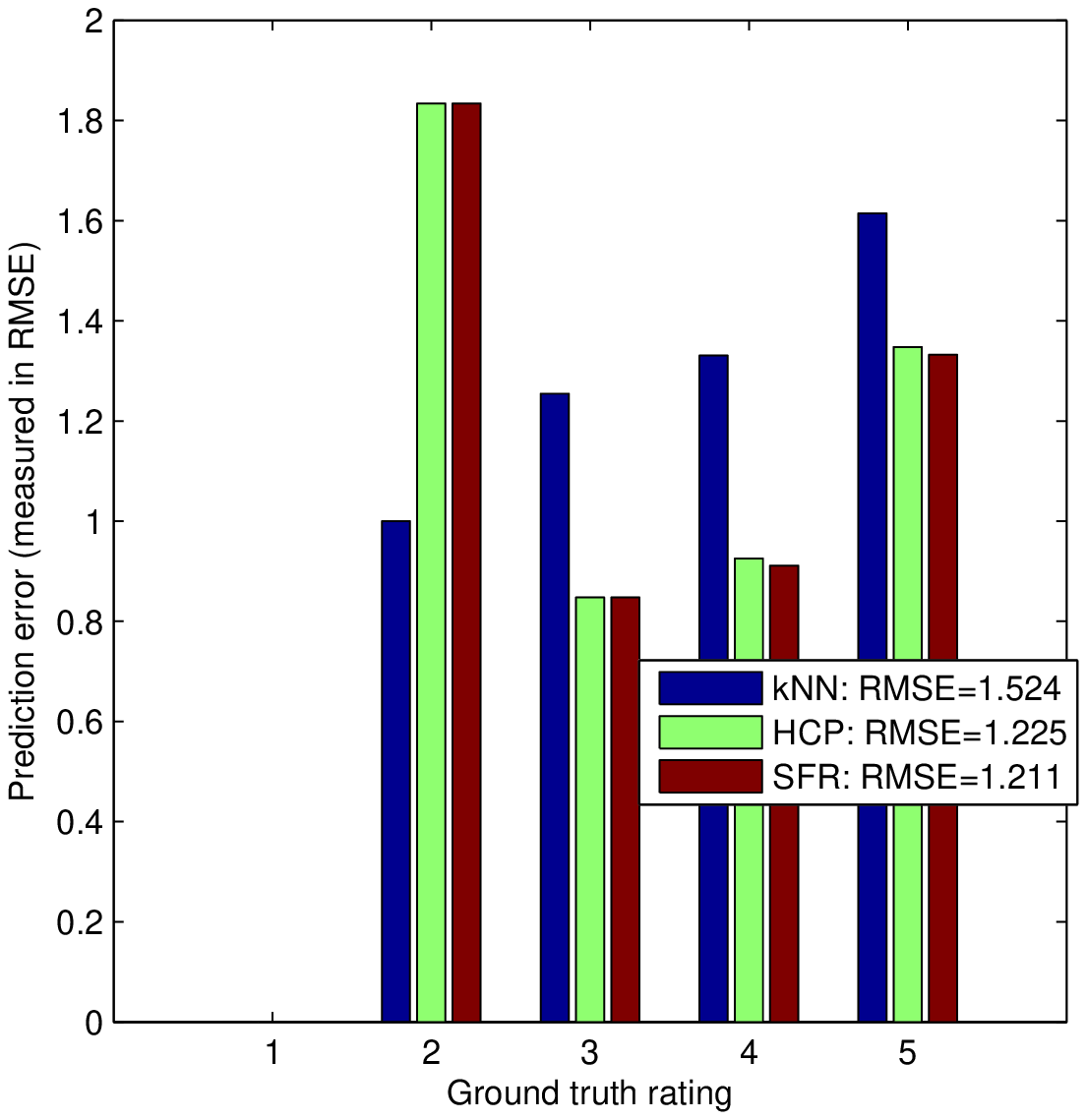}
 }
 \subfigure[MovieLens (higher)]{
    \includegraphics[scale = 0.4] {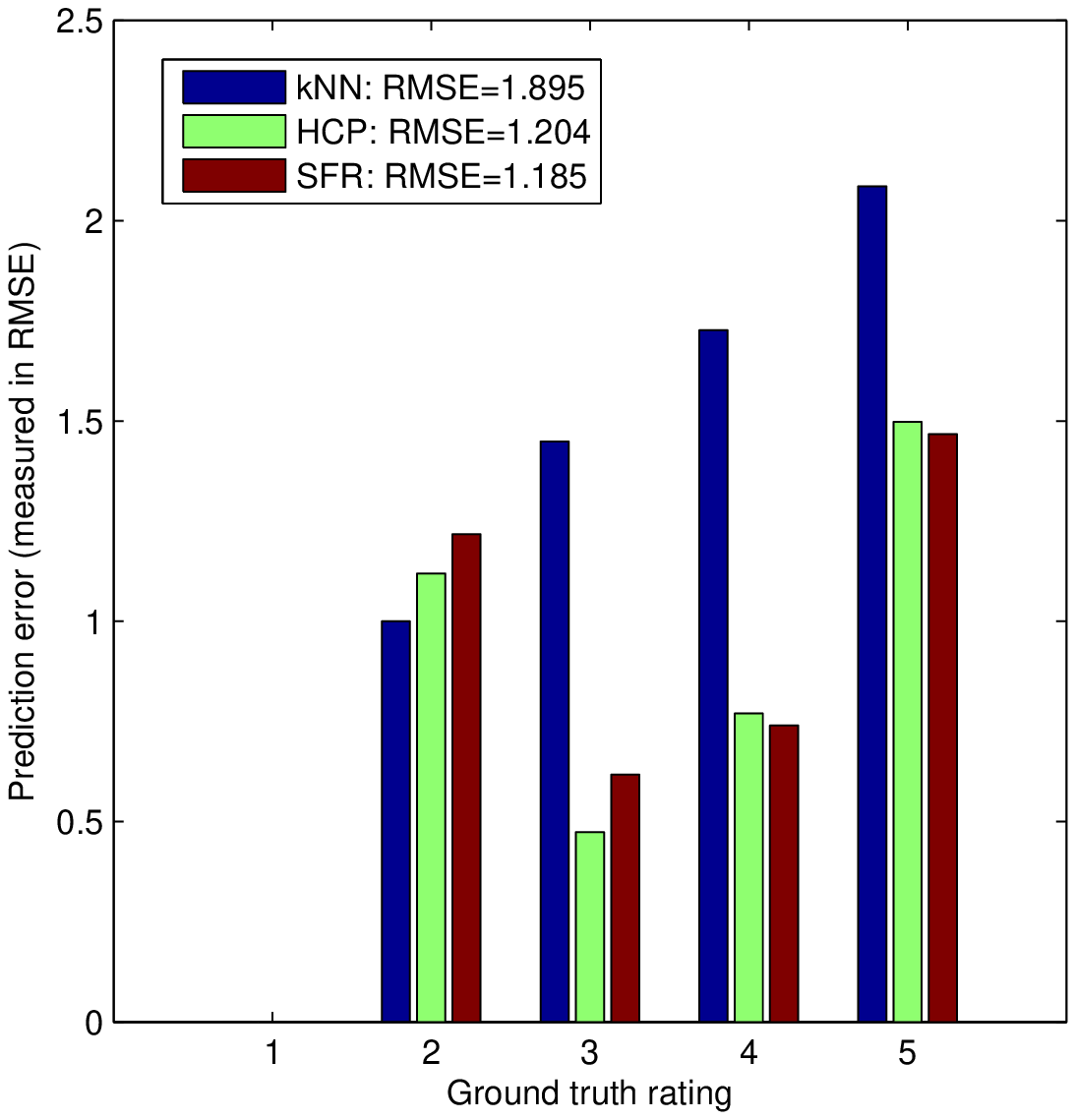}
 }
 \subfigure[Douban (lower)]{
    \includegraphics[scale = 0.4] {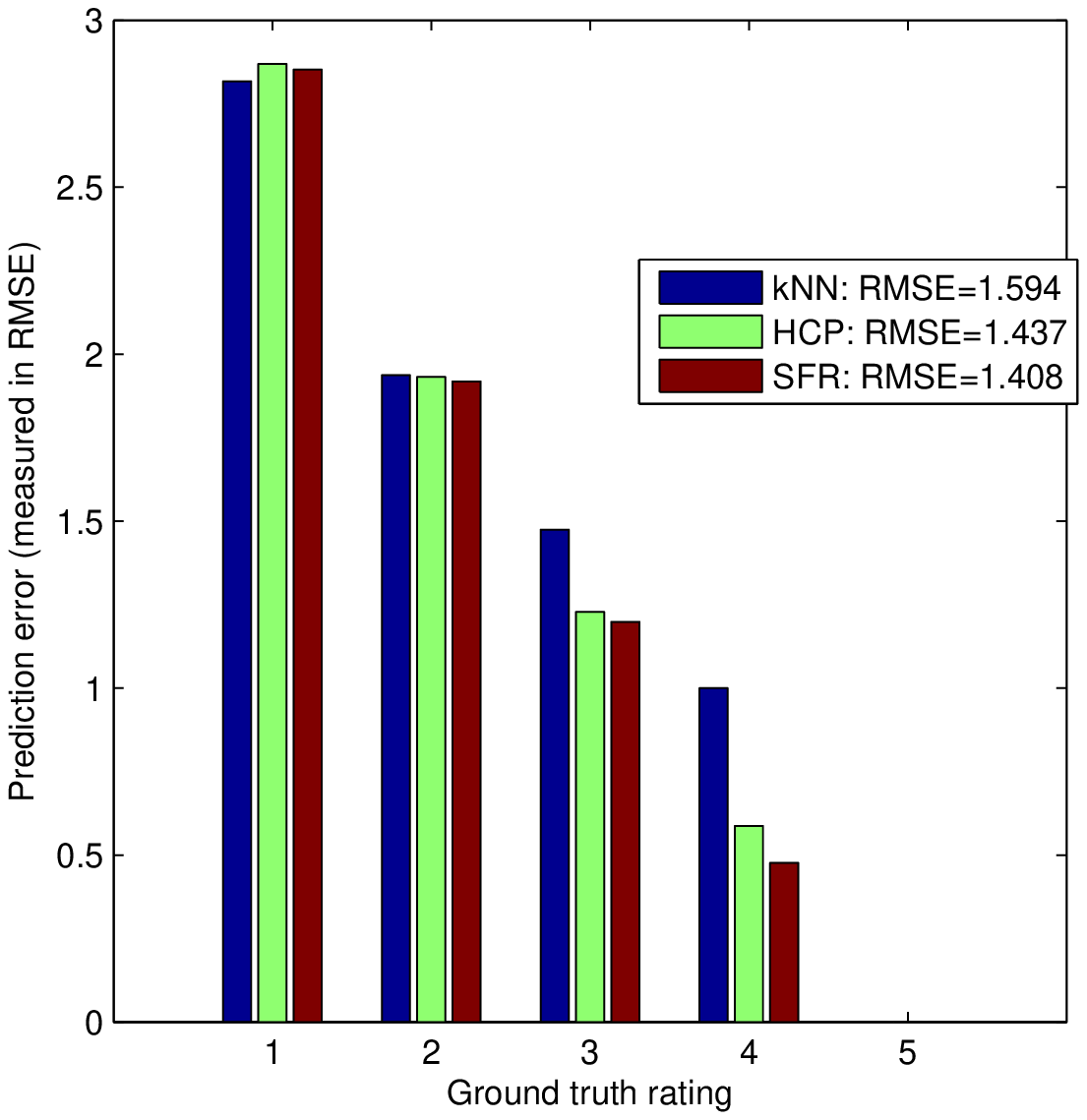}
 }
 \subfigure[Goodreads (lower)]{
    \includegraphics[scale = 0.4] {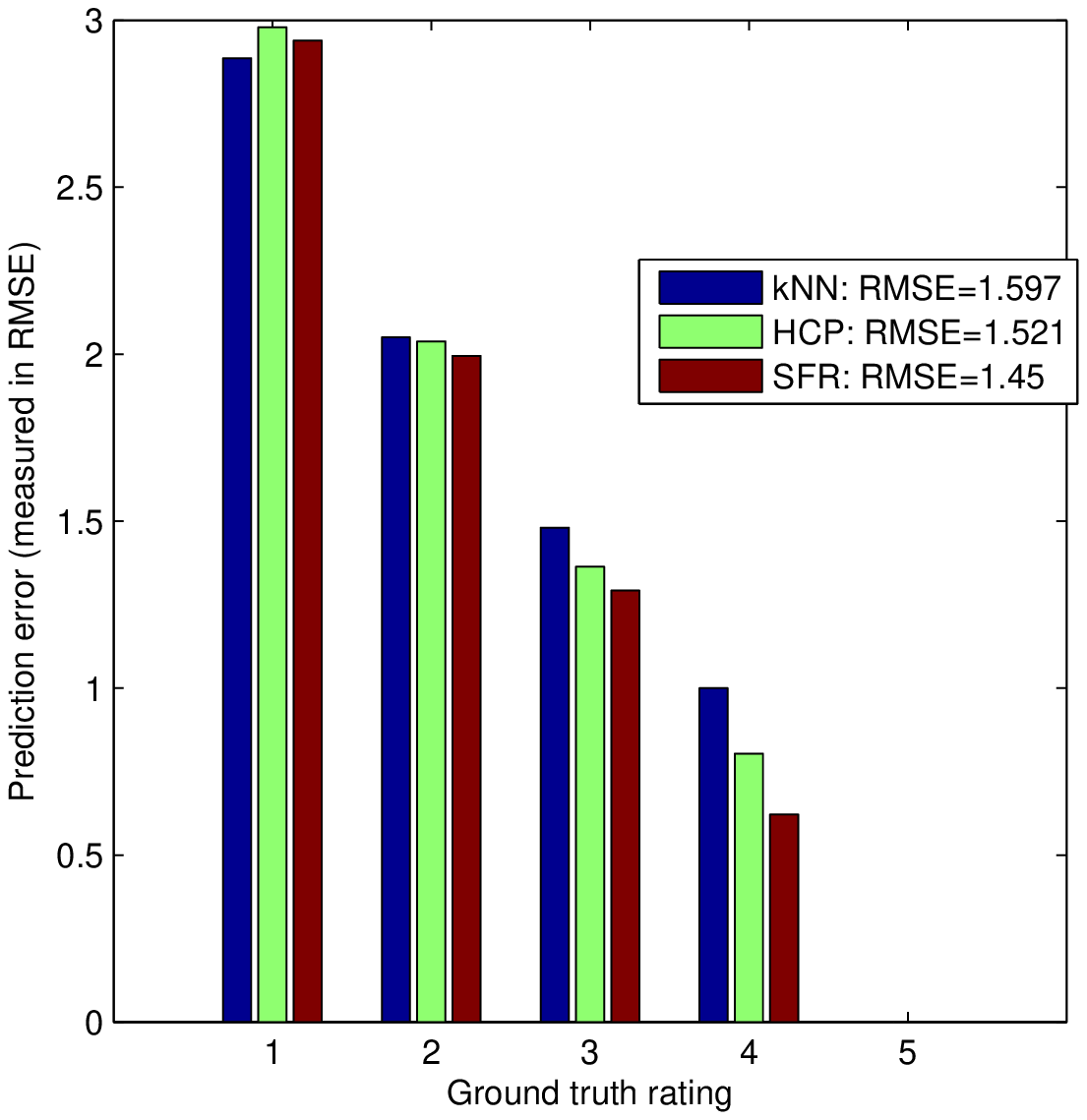}
 }
 \subfigure[MovieLens (lower)]{
    \includegraphics[scale = 0.4] {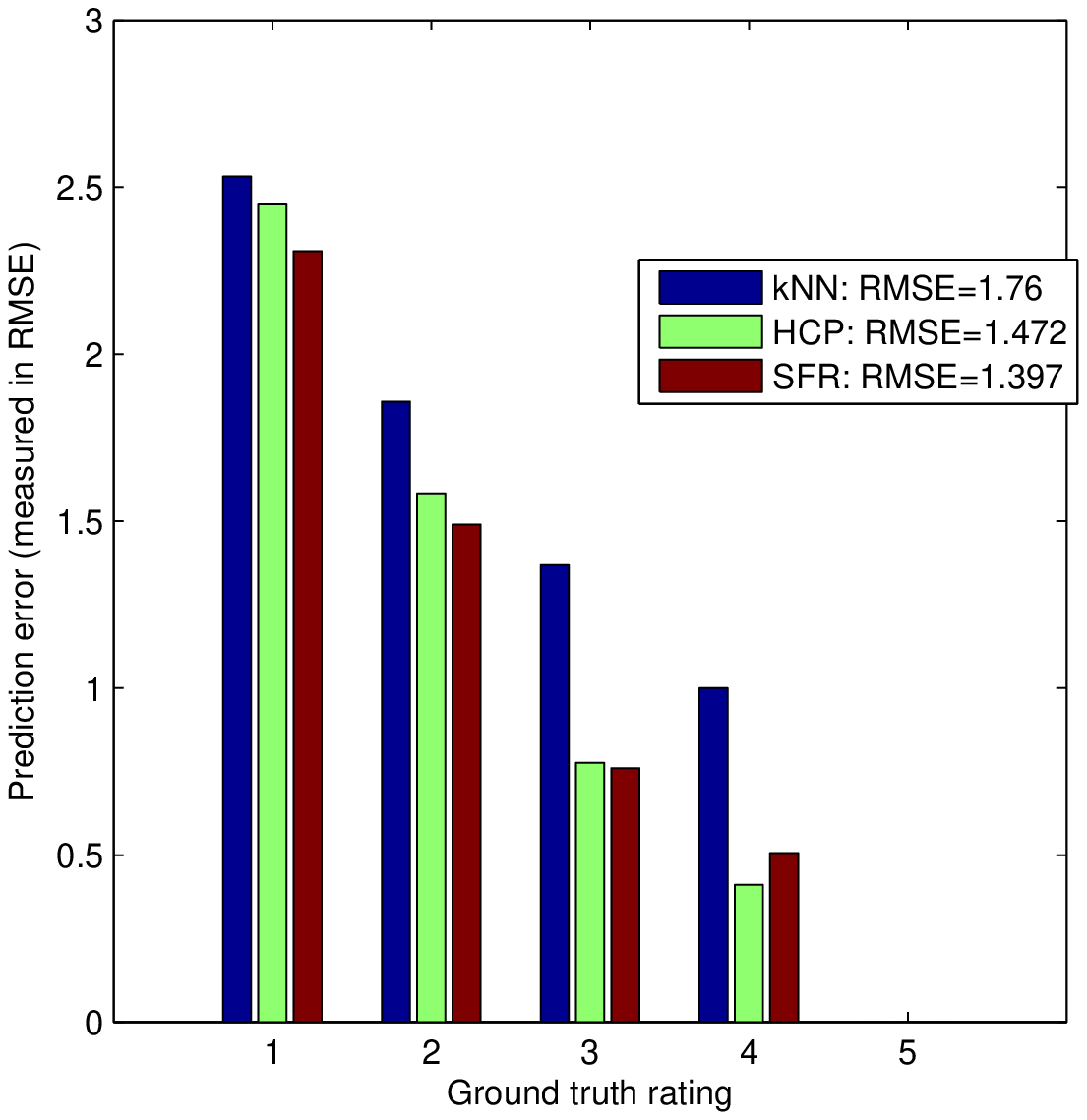}
 }
\caption{(Color online) Prediction accuracy on items which suffer from rating bound problem, measured by RMSE. Three approaches are examined to predict unobserved ratings in the testing set where the ground truth ratings are higher (top figures) or lower (bottom figures) than all neighbor observations in the training set. Our scalar function recovery approach proves to effectively reduced the prediction error in those tasks, especially when the ground truth rating is lower than all neighbor observations.}
\label{fig:result.rating.bound}
\end{figure*}

Furthermore, to analyze the approach performance on heterogeneous items, we classify the examples with the rating bound problem into subsets according to their real ratings, and report the prediction error on each subset respectively in Figure \ref{fig:result.rating.bound}. In the ``higher'' samples, both HCP and our approach largely reduce the prediction error compared with kNN, and our approach outperforms HCP slightly. The reduction is almost the same in different subsets, supporting that the advantage is consistent. The only exception is that our approach fails to outperform kNN in the subset of $2$-star rated items (higher than observed neighbors) in Goodreads dataset, where the subset contains only $6$ examples and is not statistically significant. In the ``lower'' samples, the improvement is not so large but our approach still outperforms two baselines in most cases. The improvement is more significant in high rated examples.

The worst performance of kNN attributes to the way it estimates the second derivative with partially observed neighbor ratings, while the significant improvement by HCP and our approach reveals the necessity to estimate the second derivative with all neighbors. Besides, allowing an unobserved rating not necessarily satisfying Equation (\ref{equ:2nd-derivative-vanish}) makes it possible for our approach to discover uncovered interest centers for each user, which results in the reduction of prediction error our approach achieves compared with HCP predictions. To summarize, the empirical results support our claim that our approach has the ability to gain better prediction when encountering the rating bound problem, and therefore might be more applicable to discover a user's interest centers.


%
%

\section{Related works}\label{section:backgrounds}

Personalized recommender systems have been a hot topic in the research literature for a decade \cite{collaborative-filtering-survey, collaborative-filtering-survey-2, collaborative-filtering-survey-3}.

One of the earliest personalized approaches is named content-based method, which builds a profile for each user and each book with a vector of weights on different words. The approach estimates an unobserved rating with the dot product of a user vector and a book vector \cite{content-based}. Although the approach is seldom used and soon replaced by collaborative filtering methods because its usage is limited in scenarios where an item can be explicitly parsed like a book, the form of dot product of a user vector and an item vector leaves the possibility that latent factor models arise.

Due to the advantage that a content-based method solves half of the cold start problem by building a profile for each incoming book, recently researchers are seeking for an variation of it to reduce the pain of cold start. Tag-based recommender systems are then developed to leverage user-generated tags to represent an item like videos or music which was difficult to explicitly parse \cite{tag, tag-2}.

kNN collaborative filtering techniques are developed to solve the problem left by content-based methods that items are not explicitly parsable. The series of approaches are built on the so-called similarity assumption that if two users behave similarly in the past they will behave similarly in the future \cite{collaborative-filtering}. A user-based collaborative filtering technique calculates the similarity between two users with Pearson's correlation coefficients and estimates a user's unobserved rating on a certain item with the weighted average rating that similar users post on that item. An item-based version calculates item similarity and predict in a symmetric way \cite{item-based-collaborative-filtering, item-based-collaborative-filtering-2, collaborative-filtering-empirical-study}. Recently researchers attempt to combine the two versions by simultaneously consider user-user and item-item similarities, and claim to gain better accuracy \cite{user-item-based-collaborative-filtering}. Pointing out the shortcoming that a kNN collaborative filtering technique calculates the weighted average among observed neighbor ratings which may lead to inconsistency, Zhang et al. propose the heat conduction process to estimate an unobserved rating with the weighted average rating of all neighbor ratings, no matter observed or unobserved. By simultaneously solving all unobserved ratings with a Green function, the method outperforms conventional kNN collaborative filtering methods in prediction accuracy \cite{heat-conduction}. A recent research work seeks for a unique set of global neighbors to be shared with all users and attempts to minimize the set size to achieve better accuracy and coverage \cite{Boumaza:2012:NGN:2330163.2330214}.

Latent factor models represent a user and an item with a vector in a latent feature space, and estimate a rating with the dot product of a user vector and an item vector \cite{matrix-factoriaztion-1}. Different from standard matrix factorization tasks that the whole target matrix is observed, in a recommender system only partial entries in the rating matrix are observed, and therefore controlling the risk of over-fitting becomes a key point in the inference of a latent factor model. Srebro et al. proposes a maximum margin constrain to control the structural risk represented by matrix rank \cite{maximum-margin-matrix-factorization}, which is later incorporating with compressed sensing to find an accurate completion \cite{matrix-factoriaztion-cs}. Salakhutdinov et. al introduces the probabilistic graphical model and controls the user vectors and item vectors with a Gaussian prior \cite{probabilistic-matrix-factorization, bayesian-probabilistic-matrix-factorization}. Koren controls the $l_2$-norm of user and item vectors as a regularization term in the loss function when fitting the observations \cite{svd-plus-plus, matrix-factoriaztion-survey}.

Due to the different advantages and disadvantages of the two major schools of kNN collaborative filtering and latent factor models, many researchers attempt to combine them to train a hybrid or ensemble model. A typical solution is incorporating one as a regularization into the other's framework \cite{hybrid-mf-cf-1, hybrid-mf-cf-2}.

Resource projection passes user interests back and forth in a bipartite consisting of users and items, resulting in adaptive balance between accuracy and diversity and showing good scalability when calculating on huge data \cite{resource-projection}. Restricted Boltzmann machine introduces a graphical model with a hidden layer to train user and movie profiles in an efficient and scalable way \cite{restricted-boltzmann-machine, restricted-boltzmann-machine-2}.

With the recent explosion of social networking services, researchers attempt to incorporate recommender systems with social relations to reduce the pain of sparsity and cold start problem, such as running the PageRank algorithm on a social network \cite{social-2, social-3}, analyzing social influence in recommender systems \cite{Huang:2010, Huang:2012:ESI:2124295.2124365}, or training a latent factor model to fit observed ratings and social networks simultaneously \cite{social-1, social-4, social-5}.

\section{Conclusion}\label{section:conclusion}

In recommender systems, it is crucial to accurately discover a user's interest centers, i.e., the most favorite items. Unfortunately, the commonly used neighborhood-based collaborative filtering methods fail to do so, as they suffer from the so-called rating bound problem. That is to say, the rating predicted with these methods on an item is fully bounded by those of observed ratings on neighboring items. As an interest center usually has a rating higher than the ratings of its observed neighbors, the aforementioned methods cannot accurately predict its rating at all. To overcome this significant problem, we formulated information recommendation as a problem of recovering a scalar rating function, which was further solved by optimizing the $l_\frac{1}{2}$-norm of its second derivative. Through carefully designed experiments on three real-world datasets, namely, Douban, Goodreads and MovieLens, we validated the effectiveness of the proposed approach. Specially, we found that our approach can significantly reduce the prediction error by $37\%$ when discovering interest centers, as compared to the well-known kNN collaborative filtering technique.

In the view of this paper a scalar function is defined for each user independently. Encouraged by some recent research works focusing on combining the traditional user-based and item-based collaborative filtering techniques, it is interesting to explore in our framework how to relate the scalar functions of similar users so that they could mutually borrow support on unobserved items. Besides, it is an open question to introduce a background distribution as a prior to a scalar function, e.g., the global opinion or topological property on an item might indicate its prior probability to be a source node.



\balancecolumns
\end{document}